\documentclass[useAMS,usenatbib,usegraphicx]{mn2e}
\usepackage{natbib,graphicx}
\newcounter{qcounter}

\newcommand{\msunyr}{\ensuremath{\mathit{M}_{\odot}{\rm yr}^{-1}}}   
\newcommand{\kms}{\ensuremath{{\rm km\,s^{-1}}}}                   
\newcommand{\K}{\mathrm{K}}
\newcommand{\lsun}{\ensuremath{\mathit{L}_{\odot}}}                  
\newcommand{\rsun}{\ensuremath{\mathit{R}_{\odot}}}                  

\newcommand{\lstar}{\ensuremath{\mathit{L}_{\star}}}                 
\newcommand{\mdot}{\ensuremath{\dot{M}}}                             
\newcommand{\rstar}{\ensuremath{\mathit{R}_{\star}}}                 
\newcommand{\teff}{\ensuremath{\mathit{T}_{\rm eff}}}                
\newcommand{\vinf}{\ensuremath{v_{\infty}}}                          
\newcommand{\tstar}{\ensuremath{\mathit{T}_{\star}}}                 

\newcommand{\ang}{\ensuremath{\mathrm{{\AA}}}}                
\newcommand{\phisp}{\ensuremath{\phi_{\mathrm{sp}}}}                 
\newcommand{\phiorb}{\ensuremath{\phi_{\mathrm{orb}}}}                 
\newcommand{\tauross}{\ensuremath{\tau_{\mathrm{Ross}}}}                 
\newcommand{\etaa}{\ensuremath{\eta_{\mathrm{A}}}}                 
\newcommand{\etab}{\ensuremath{\eta_{\mathrm{B}}}}                 

\newcommand\ion[2]{#1$\;${\scshape{#2}}}

\newcommand{\hststis}{{\it HST}/STIS}                          

\title[On the influence of the companion star in Eta Carinae ]{On the influence of the companion star in Eta Carinae: 2-D radiative transfer modeling of the ultraviolet and optical spectra\thanks{Based on observations made with the {\it Hubble Space Telescope} Imaging Spectrograph under programs 9420 and 9973.}}
\author[J. H. Groh et al.]{Jose H. Groh$^1$, D. John Hillier$^2$, Thomas I. Madura$^1$, and Gerd Weigelt$^1$\\
$^1$Max-Planck-Institut f\"ur Radioastronomie, Auf dem H\"ugel 69, D-53121 Bonn, Germany\\
$^2$Department of Physics and Astronomy, University of Pittsburgh, 3941 O'Hara Street, Pittsburgh, PA, 15260, USA}

\begin{document}

\date{Accepted 23 March 2012 . Received 23 January 2012}

\pagerange{\pageref{firstpage}--\pageref{lastpage}} \pubyear{2012}

\maketitle

\label{firstpage}

\begin{abstract}

We present two-dimensional (2-D) radiative transfer modeling of the Eta Carinae binary system accounting for the presence of a wind-wind collision (WWC) cavity carved in the optically-thick wind of the primary star. By comparing synthetic line profiles with spectra obtained with the {\it Hubble Space Telescope}/Space Telescope Imaging Spectrograph near apastron, we show that the WWC cavity has a strong influence on multi-wavelength diagnostics. This influence is regulated by the modification of the optical depth in the continuum and spectral lines. We find that H$\alpha$, H$\beta$, and \ion{Fe}{ii} lines are the most affected by the WWC cavity, since they form over a large volume of  the stellar wind of the primary. These spectral lines depend on latitude and azimuth since, according to the orientation of the cavity, different velocity regions of a spectral line are affected. For 2-D models with orientation corresponding to orbital inclination angle $110\degr \la i \la  140\degr $ and longitude of periastron $210\degr \la \omega \la 330\degr$, the blueshifted and zero-velocity regions of the line profiles are the most affected by the cavity. These orbital orientations are required to simultaneously fit the UV and optical spectrum of Eta Car around apastron, for a half-opening angle of the cavity in the range 50\degr--70\degr. We find that the excess P-Cygni absorption seen in H$\alpha$, H$\beta$, and optical \ion{Fe}{ii} lines in 1-D spherical models becomes much weaker or absent in the 2-D cavity models, in agreement with the observations. The observed UV spectrum of Eta Car is strongly dominated by absorption of \ion{Fe}{ii} lines that are superbly reproduced by our 2-D models when the presence of the low-density WWC cavity is taken into account. Small discrepancies still remain, as the P-Cygni absorption of H$\gamma$ and H$\delta$ is overestimated by our 2-D models at apastron. We suggest that  photoionization of the wind of the primary by the hot companion star is responsible for the weak absorption seen in these lines. Our CMFGEN models indicate that the primary star has a mass-loss rate of $8.5\times10^{-4}~\msunyr$ and wind terminal velocity of $420~\kms$ around the 2000--2001 apastron.
\end{abstract}


\section{Introduction}  \label{intro}

\defcitealias{bh05}{BH05}
\defcitealias{hillier06}{H06}
\defcitealias{hillier01}{H01}

The evolution and fate of massive stars is greatly affected by strong mass loss that occurs via stellar winds throughout the star's lifetime (e. g., \citealt{conti76,langer94,meynet03}). Rare giant eruptions of several solar masses may also occur during the unstable Luminous Blue Variable phase \citep[LBV;][]{hd94,so06}, but the frequency and total mass lost via this phenomenon is uncertain. To add further uncertainty, the interaction with a binary companion has been suggested as a possible trigger to giant eruptions \citep{kashi10,pastorello10,smith11}, which highlights the concept that single and binary massive stars may have different mass loss histories and fates. Studies of the most massive members of the LBV class, such as Eta Carinae, are needed to gain insights into the mass loss processes in massive stars.

Even though it is one of the most observed and scrutinized objects in the sky, the superluminous Eta Carinae still poses a great challenge to understanding the fate of massive stars. Located in the Trumpler 16 cluster in the Carina nebula and at a distance of $2.3\pm0.1$~kpc \citep{walborn73,allen93,smith06}, Eta Car is composed of a bright central object whose luminosity ($L_{\star}\geq 5\times 10^6~\lsun$, \citealt{dh97,smith03b}) ranks among the top in the Galaxy. First proposed by \citet{damineli97}, it is now widely accepted that Eta Car is a massive, colliding-wind binary system with a high eccentricity ($e\sim0.9$, \citealt{corcoran05}) and an orbital period of $2022.7 \pm 1.3$ d \citep{damineli08_period}. This scenario is supported by multi-wavelength observations in X-rays \citep{corcoran97,corcoran01,corcoran10,pittard98,pc02,ishibashi99,corcoran05,hamaguchi07,henley08}, ultraviolet \citep{smith04,iping05,mg12}, optical \citep{vg03,vg06,lajus03,lajus09, lajus10, sd04,nielsen07,damineli00,damineli08_multi,damineli08_period,teodoro12,mehner11}, near-infrared \citep{feast01,whitelock04,gull09,gmo10,gnd10}, and radio wavelengths \citep{duncan03,abraham05a}. 

The orientation of the orbit has been the subject of debate over the last decade.Most authors agree on a longitude of periastron of $\omega\simeq240\degr-270\degr$, as proposed initially by \citet{damineli97}, while others have found $\omega \simeq50\degr-90\degr$ (e.g., \citealt{abraham05a,kashi08a}). Recently, \citet{madura11} determined the orbital orientation for the first time in three-dimensional (3-D) space. Based on velocity- and spatially-resolved [\ion{Fe}{iii}] emission, these authors found that the Eta Car system has an orbital inclination angle $i \approx 130^{\circ}$ to $145^{\circ}$, longitude of periastron $\omega \approx 240^{\circ}$ to $285^{\circ}$, and an orbital axis projected on the sky at a position angle $\mathrm{PA}_{z} \approx 302^{\circ}$ to $327^{\circ}$ from North to East. These values are supported by a multitude of multi-wavelength observations \citep{okazaki08,parkin09,parkin11,gnd10,gull11,mg12}.

The primary star (hereafter \etaa) has been shown to be an LBV, with a high mass-loss rate of $\sim10^{-3}~\msunyr$, wind terminal velocity of $\sim500~\kms$, and dominates the luminosity of the system (\citealt{hillier01}, hereafter H01). Line profiles of H$\alpha$, H$\beta$, and \ion{Fe}{ii}, obtained at different positions in the Homunculus nebula that surrounds the binary system, support a latitude dependent wind only around apastron, with the wind becoming more spherical during periastron \citep{smith03}. A scenario with a single, rapid-rotating $\etaa$ causing line profile variations as a function of latitude has been proposed to explain the deeper P-Cygni absorption profiles and faster wind velocities observed at the pole \citep {smith03}. Rapid rotation may also explain the elongated K-band photosphere directly resolved with interferometric measurements \citep{vb03,weigelt07}. To explain these observations in a single-star scenario,  \citet{gmo10} found that \etaa\ must rotate between 77\% to 92\% of the critical velocity for break-up, with its rotational axis likely misaligned with the Homunculus polar axis.

The nature and evolutionary state of the companion (hereafter \etab), however, are far less certain since it has never been directly observed. This is because, in the visible and longer wavelengths, \etaa\ is a few orders of magnitude brighter than \etab\ \citep{hillier06}. X-ray observations suggest a  wind terminal velocity of $\sim3000~\kms$ for \etab\ \citep{pc02,okazaki08,parkin09,parkin11}, which is in line with the detection of high-velocity material ($\sim2000~\kms$) coming from the wind-wind collision (WWC) zone \citep{gnd10,henley08}. Studies of the ionization effects of the companion on the nearby ejecta have yielded constraints on its temperature ($\teff\simeq36,000-41,000~\K$, \citealt{mehner10,verner05,teodoro08}) and luminosity ($10^5 \lsun \la \lstar \la 10^6 \lsun$, \citealt{mehner10}).

The presence of a close, hot companion can potentially affect the integrated spectrum of Eta Car in two main ways. First, \etab, depending on its luminosity and temperature, may photoionize part of the wind of \etaa, affecting H, \ion{He}{i}, and \ion{Fe}{ii} lines, among others. This photoionization effect has been qualitatively discussed in various contexts \citep[e.\,g.][]{damineli97,damineli08_multi,hillier01,smith03,davidson05,nielsen07,gull09,richardson10,madura11}. 

Second, 3-D hydrodynamical simulations have shown that the wind of \etab\ greatly affects the geometry of the wind of \etaa, creating a low-density WWC cavity in the wind of \etaa\ and a thin, dense wind-wind interacting region between the two winds \citep{pc02,okazaki08,parkin09,parkin11,madura11}. The effects of this WWC cavity on the spectrum of Eta Car have been qualitatively discussed in the literature, as a quantitative study required yet unavailable 2-D radiative transfer models. \citetalias{hillier01} proposed that different lines will be affected differently by \etab\ depending on the size of their formation region. \citet{richardson10} suggested that the carving of the wind of \etab\ might be responsible for explaining variations in the H$\alpha$ emission seen during periastron. \citet{teodoro12} suggested that H$\delta$ emission is not affected by the WWC during most of the orbit, while the H$\alpha$ and \ion{Fe}{ii} emissions are. 

Our goal in this paper is to perform a quantitative investigation of the influence of \etab\ on the spectrum of \etaa, particularly due to the presence of a cavity in the wind of \etaa. If these effects are significant, a detailed spectroscopic analysis might constrain the orbital parameters of the binary system and the individual stellar and wind parameters. Our ultimate goal is to gain insights on the nature of both stars. For instance, is $\etaa$ a rapid rotator, or can the presence of a close companion equally explain the available observations? \citet{gmo10}, based on the 2-D models discussed here, showed that the aforementioned $K$-band interferometric observations of Eta Car can be explained by both scenarios.

With these goals in mind, we apply, for the first time, 2-D radiative transfer models to analyze multi-wavelength spectral regions of Eta Car, from the ultraviolet (UV) to the near-infrared. Here, we focus on the analysis of high spatial resolution spectroscopic observations obtained with the {\it Hubble Space Telescope}/Space Telescope Imaging Spectrograph ({\it HST}/STIS) at UV and optical wavelengths around apastron. At this orbital phase, time-dependent and 3-D effects due to orbital motion are minimized. 

We briefly describe in Sect. \ref{obs} the archival spectroscopic data of Eta Car used in this analysis. Section \ref{spherical} presents a revised 1-D radiative transfer model of \etaa\ using the code CMFGEN \citep{hm98}. Section \ref{cavitymodel} describes our 2-D radiative transfer model for the wind of \etaa\ including the WWC cavity and a dense WWC zone. In  Section \ref{standardparamsline} we analyze how the presence of the cavity affects spectral lines of \etaa\ during apastron. Section \ref{compareobs} presents a detailed comparison between the 2-D model spectrum and the observed spectrum of the central source in Eta Car around apastron. We examine  in Sect. \ref{orbstudy} the influence of, and constraints to, the orbital parameters based on the analysis of the UV and optical spectra. The effects of the WWC cavity properties, such as the half-opening angle, are studied in Sect. \ref{paramstudy}. Effects of the WWC cavity on the continuum are discussed in Sect. \ref{standardparamscont}. A discussion on the ionization effects expected from the hot companion is presented in Sect. \ref{ionization} and our concluding remarks are presented in Sect. \ref{conc}.

In forthcoming papers, we will investigate variations as a function of orbital phase, and the latitudinal and azimuthal dependencies of the spectrum that can be probed using reflected spectra off the ejecta around Eta Car.

\section{Observations} \label{obs}

\begin{table}
\caption{\label{tab1} \hststis\ Eta Car spectroscopic observations used in this paper} 
\centering
\begin{tabular}{l c c c}
\hline\hline
Date & $\phisp$ & JD &Wavelength Range\\
        &              &           &   (\AA) \\
\hline
2000 Mar 23 & 10.410 & 2451626 & 1170--1700\\
2001 Apr 17 & 10.603 & 2452017 & 1640--10100 \\
\hline
\end{tabular}
\end{table}

We employ UV and optical spectra of Eta Car comprising the inner $\sim0\farcs2$, minimizing the amount of contamination of the central source spectrum by the nearby ejecta. Seeing-limited ground-based observations are not adequate for our purposes as they are mainly contaminated by 1) bright, narrow forbidden and permitted line emission from the ejecta \citep{weigelt86,weigelt95,davidson95}, and 2) faint, extended ($\sim0.3-1.0\arcsec$) high-ionization forbidden line emission that arises in the wind-wind interacting region \citep{gull09,gull11,madura11}. Both of these features, which are likely powered by ionizing photons from \etab\ \citep{verner05,mehner10,madura11}, are not modeled in this paper. 

The archival UV and optical  \hststis\ observations analyzed in this paper were obtained by the {\it HST} Eta Car Treasury project\footnote{The reduced data are available online at http://etacar.umn.edu}. A large wavelength coverage is needed to perform the detailed spectroscopic analysis, and reach the goals of this paper. Therefore, we use spectra gathered near apastron at $\phisp=10.410$ (2000 Mar 23) and $\phisp=10.603$ (2001 Apr 17), where $\phisp$ corresponds to a given phase of the spectroscopic cycle of Eta Car. We adopt the ephemeris from \citet{damineli08_period} and the cycle labeling from \citet{gd04}.

\begin{figure*}
\resizebox{\hsize}{!}{\includegraphics{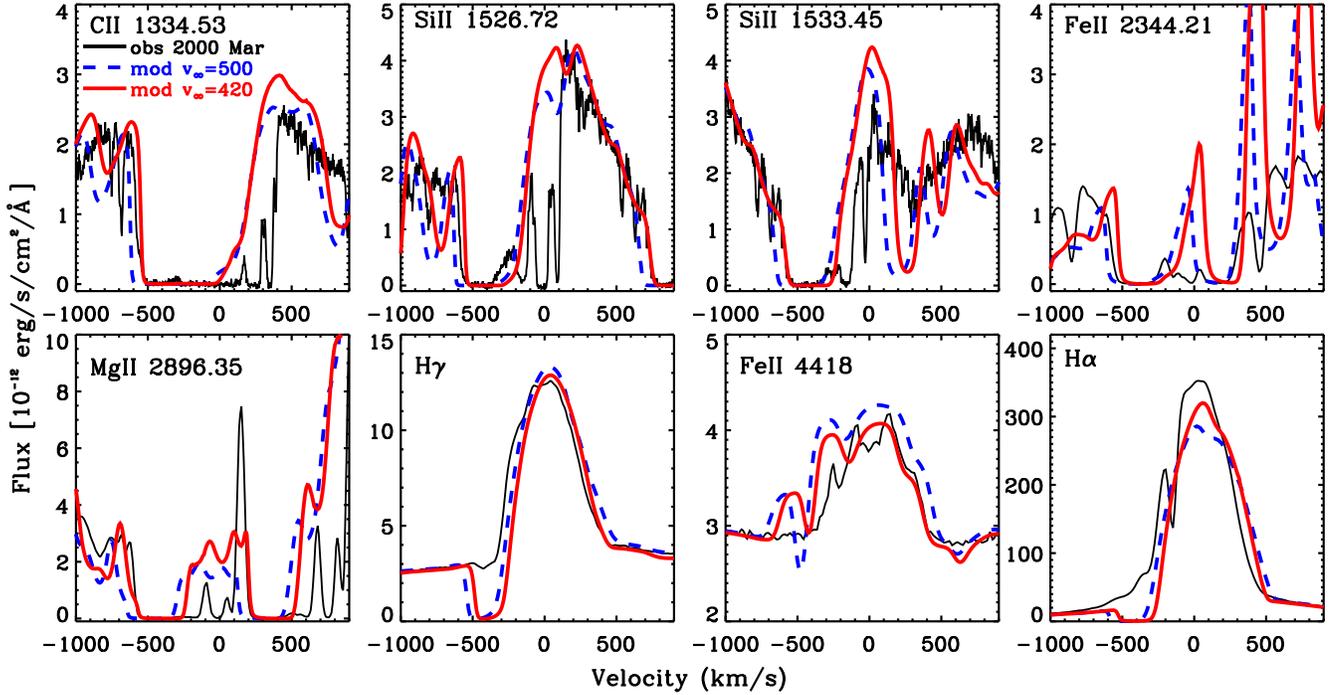}}
\caption{\label{comparevinf}{Spectrum observed with \hststis\ around apastron (black solid line) compared with the spherical CMFGEN model from \citetalias{hillier01} with $\vinf=500~\kms$ (blue dashed) and our revised spherical CMFGEN model with $\vinf=420~\kms$ (red solid).  }}
\end{figure*}

Ultraviolet spectroscopic observations in the range 1170--1700~{\AA} were obtained with the MAMA echelle grating E140M, providing a spectral resolving power of $R\sim48000$ throughout this spectral region using an aperture of $0\farcs2 \times 0\farcs2$. Following previous studies \citep{nielsen05,gnd10},  a 7-pixel (0\farcs0875) extraction centered on the central source was adopted to minimize contamination from the extended ejecta. Custom IDL routines developed by one of us (JHG), assuming a boxcar function, were used to extract the spectrum.  Optical and near-UV observations were gathered with the $52\arcsec \times 0\farcs1$ aperture, employing the G430M and G750M gratings to cover the spectral region 1640--10100~{\AA} with $R\sim8000$. An aperture of 6 half-pixels (0\farcs152) was used for the spectral extraction with custom IDL routines. We refer the reader to \citet{smith03}, \citet{davidson05}, \citet{hillier06}, \citet{nielsen07}, and \citet{gull09} for further details about the observations and data reduction. Table \ref{tab1} summarizes the \hststis\ data used in this paper.

\section{The  1-D spherical radiative transfer model of \etaa} \label{spherical}

Since our 2-D models (Sect. \ref{cavitymodel}) are based on the spherically-symmetric models of \etaa\ \citepalias{hillier01,hillier06}, we briefly summarize here the relevant model parameters obtained using the non-LTE, fully line blanketed radiative transfer code {\sc CMFGEN} \citep{hm98}. 

We use similar diagnostics and CMFGEN models as those discussed by \citetalias{hillier06}, with the key difference being a lower wind terminal velocity of $\vinf=420~\kms$ and correspondingly lower mass-loss rate of $\mdot = 8.5 \times 10^{-4}~\msunyr$ (Sect. \ref{lowervinf}). Because of the lower $\vinf$, several other parameters need to be adjusted to match the observed spectra. The parameters of our revised 1-D CMFGEN model are shown in Table \ref{params}. We note that, as in \citetalias{hillier01}, models with $\rstar$ in the range $60-480$ \rsun\ produce similar fits to the observed spectrum. The atomic model and abundances are the same as described in \citetalias{hillier01} and \citetalias{hillier06}, with the exception that we adopt  a solar Fe abundance from \citet{asplund05}.

\subsection{Discrepancies when fitting the observed spectrum} \label{sphericaldiscrepancies}

Although the spherical model of \etaa\ reproduces reasonably well the emission lines seen in the \hststis\ spectrum, some discrepancies remain (see Sect. 10 of \citetalias{hillier01}).  Among these, the spherical model of Eta Car is not able to fit the UV and optical spectra simultaneously (Fig. 12 of \citealt{hillier01}). The revised CMFGEN models presented here have the same difficulties that we outline below (Fig. \ref{comparevinf}).

Significant improvements to the fitting of the UV spectrum were obtained when including a ``coronagraph" that blocks the light of the inner 0\farcs033 region \citep{hillier06}. In addition, the CMFGEN model also has difficulties with reproducing the central intensity of both the high Balmer lines and H$\alpha$ simultaneously \citepalias{hillier01}. The optical \ion{He}{i} lines are blueshifted compared to the 1-D CMFGEN model, which argues against them being formed in the wind of \etaa\ \citep{nielsen07}.

The spherical model also overestimates the amount of absorption in the UV and optical lines, in particular in H and Fe lines (Figs. 5, 6, and 7 of \citealt{hillier01}). This is surprising in a single star scenario, since CMFGEN models are able to reproduce the H and \ion{Fe}{ii} P-Cygni profiles  of HDE 326825, which has an optical spectrum similar to that of \etaa\  \citep{hillier98}.  In \citetalias{hillier01}, the spherical model of \etaa\ was compared to observations obtained just after periastron ($\phisp=10.041$). Since the amount of observed P-Cygni absorption is even lower around apastron \citep{damineli98,davidson05,nielsen07,damineli08_multi}, the quality of the fits to the absorption profiles becomes much worse when comparing to the apastron data (Sect. \ref{standardparamsline}). The fits to the P-Cygni absorption component and UV spectrum could be improved by significantly decreasing $\mdot$ in the models, which conversely makes the fit to the emission lines much worse  \citep{hillier06}. 

\subsection{A lower wind terminal velocity for \etaa} \label{lowervinf}

When compared to \hststis\ observations obtained at apastron, we find that the original CMFGEN spherical model of \etaa\ with $\vinf = 500~\kms$ overestimates the amount of emission on the red side of most optical lines, such as H and \ion{Fe}{ii} lines (Fig.~\ref{comparevinf}). In addition, that model overestimates the maximum velocity of the P-Cygni absorption component of low-ionization resonance lines, such as \ion{C}{ii} $\lambda$1334--1335, \ion{Si}{ii} $\lambda$1526--1533, \ion{Mg}{ii} $\lambda$2796--2802, and many resonance-like \ion{Fe}{ii} lines around $\lambda$2400 and $\lambda$2600 (Fig.~\ref{comparevinf}).

We find that a spherical model with a wind  terminal velocity of \etaa\ of $\vinf = 420~\kms$ provides improved fits to the UV and optical spectra.  As shown in Figure~\ref{comparevinf}, a model with this slightly lower value of $\vinf$ provides a much better fit to the red emission of optical lines of H and \ion{Fe}{ii}. Conversely, the fits to the blue side of the H emission  become slightly worse. However, we prioritize the fitting of the red side of the lines profiles with the fore-knowledge that, for our preferred orbital orientation with $i=138\degr$ and $\omega=260\degr$, the red side of the line profiles are the least affected by the presence of the WWC cavity (Sect. \ref{standardparamsline}). The blue side of the optical line profiles are extremely affected not only by the WWC cavity as modeled in this paper, but perhaps also the driving of the wind of \etaa\ might be modified by the ionizing photons of the hot, luminous \etab.

We also analyzed the behavior of the maximum velocity of black absorption (hereafter $v_\mathrm{black}$) of the low-ionization UV resonance lines of \ion{C}{ii}, \ion{Si}{ii}, \ion{Fe}{ii}, and \ion{Mg}{ii}, which has been demonstrated to be the best diagnostic to $\vinf$ \citep{prinja90}. Figure \ref{comparevinf} shows that also the low-ionization UV resonance lines are more consistent with $\vinf \simeq 420~\kms$  than $\simeq500~\kms$ (Fig. \ref{comparevinf}). Our models suggest that, at least at apastron, material with velocities above $\sim 420~\kms$ appears only on the front side of the wind of $\etaa$, and are probably caused by turbulence both inherent to radiative-driven winds \citep{owocki88} or the WWC, with possible mixing of the two winds. Similar to the 1-D models of \citetalias{hillier01}, our models assume a microturbulent velocity of $30~\kms$ that increases up to $50~\kms$ at larger distances in the wind of \etaa.

The value of $\vinf$ obtained here is slightly lower than the $\vinf=500~\kms$ proposed by \citetalias{hillier01} based on \hststis\ data obtained just after periastron ($\phisp=10.041$, 1998 Mar). The change could be real or a result of using different diagnostics to determine $\vinf$, and radiative transfer modeling near periastron is needed to properly address this issue. There are also issues related to the precise meaning of $\vinf$, such as the influence of microturbulence and non-monotonicity arising from radiation line driving instabilities. At this point, given that $v_\mathrm{black}$ of the UV resonance lines does not change as a function of $\phiorb$ \citep{hillier06,gnd10,mehner11b}, it seems more likely that $\vinf \simeq 420~\kms$ is more appropriate for material in the undisturbed wind of \etaa\ at all phases.

The value of $\vinf$ found here based on detailed spectroscopic analysis is lower than estimates using the same \hststis\ data but ad-hoc assumptions without a quantitative physical model ($\vinf=540~\kms$, \citealt{smith03}; $\vinf=600~\kms$, \citealt{mehner11b}). The higher values of $\vinf$ from previous studies illustrate the challenge of obtaining quantitative information about the velocity field of stellar winds without employing radiative transfer models.

Interestingly, the modeling of high-ionization forbidden-line emission of \etaa, using an independent 3-D radiative transfer code  and assuming  $\vinf=500~\kms$, overestimated the amount of blueshifted emission compared to the observations \citep{madura11}. These authors suggested that a lower value of $\vinf$ could possibly reconcile the 3-D models and observations, and our results for the optical and UV lines of Eta Car provide further support for a lower $\vinf$.

Despite improving the fits to the line profiles in the UV and optical spectra, the discrepancies in the comparison with the observations (Sect. \ref{sphericaldiscrepancies}) are still present. These are the main motivation for attempting a 2-D modeling of the spectrum of Eta Car, as we want to investigate whether the discrepancies found by \citetalias{hillier01} could be caused by the influence of \etab.

\section{The 2-D radiative transfer model: primary wind modified by the wind-wind collision cavity}  \label{cavitymodel}

\begin{figure*}
\center
\resizebox{0.99\hsize}{!}{\includegraphics{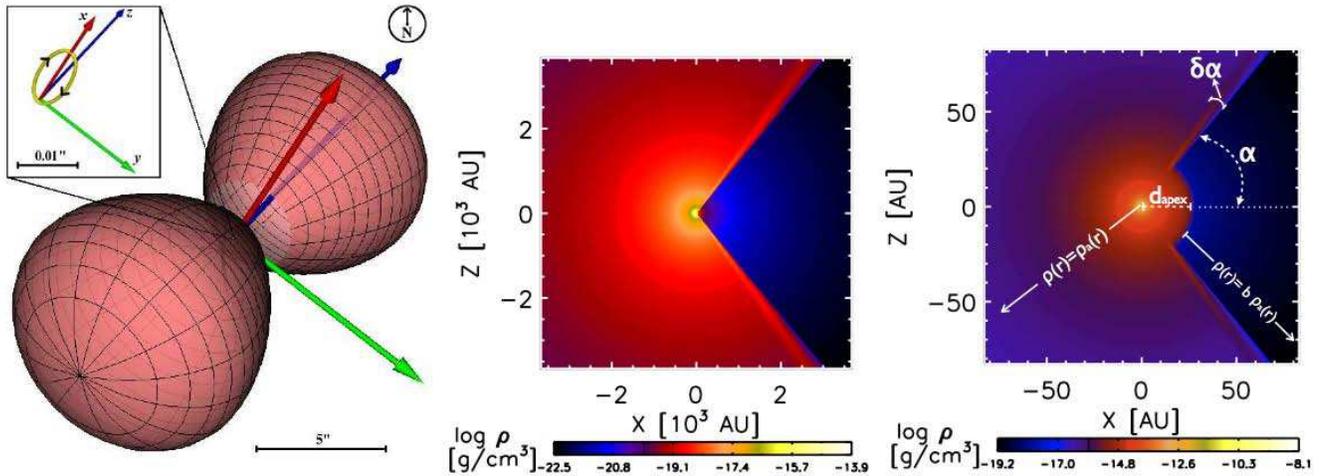}}
\caption{\label{cavity_sketch} {\it Left panel:} Geometry of the Eta Car orbit (yellow ring in upper left inset) on the sky from \citet{madura11}. The red arrow corresponds to the semi-major axis $x$, the green arrow to the semi-minor axis $y$, and the blue arrow to the orbital axis $z$.  {\it Middle and right panels:} Assumed geometry of the wind of \etaa\ modified by the WWC cavity during apastron, showing a slice in the $x-z$ plane (i.e., the binary system is viewed edge-on). The middle panel shows the large distance range covered by our 2-D radiative transfer computations ($\sim5500$~AU, $\simeq2\arcsec$), while the right panel shows a zoom into the inner regions of the WWC cavity. Both panels are color-coded to log density.}
\end{figure*}

We use an updated version of the 2-D code of \citet[hereafter BH05]{bh05} to create a low-density WWC cavity in the optically-thick, spherical wind of \etaa. Our 2-D model also includes the presence of a dense WWC zone, corresponding to the post-shocked primary wind. Figure \ref{cavity_sketch} shows the assumed geometry of the WWC cavity. We refer the reader to \citetalias{bh05} for further details on the original code and to \citet{ghd06,groh08,gdh09,gmo10} and \citet{driebe09} for details on the code's applications. Our implementation is flexible enough to be applied to other colliding-wind massive binaries, such as WR140.

The 2-D code uses as input several quantities (e.\,g., energy-level populations, ionization structure, radiation field) from the spherically-symmetric model of \etaa\ computed with {\sc CMFGEN}. The code allows for the specification of any arbitrary latitude-dependent variation of the wind density $\rho$ and $\vinf$. We initially consider a spherical wind for \etaa, with no intrinsic dependency of $\rho$ and $\vinf$ on the colatitude $\theta$. To create a low-density cavity in the wind of 
\etaa, we modify the spherical $\rho(\theta)$ so that the cavity has a conical surface with half-opening angle $\alpha$ (Fig. \ref{cavity_sketch}). Material within the cavity is assigned a density $b$ times lower than that of the spherical wind model of \etaa. Thus, the parameter $b$ reflects the different level populations and density of the wind of \etab, which fills in the cavity, with respect to the wind of \etaa. At the moment, neither the flux from \etab\ nor the post-shocked secondary-wind are included. The WWC zone is approximated by cone walls of angular thickness $\delta_\alpha$ and a density contrast of $f_\alpha$ times higher than the wind density given by the spherical model of \etaa\ at a given radius. According to mass conservation \citep{gull09},
\begin{equation}
\label{eqfalpha}
f_\alpha = [1-\cos(\alpha)]/[\sin(\alpha)\delta_\alpha]. 
\end{equation}

The simple conical shape is justified for orbital phases sufficiently away from periastron, when the cavity has an approximately 2-D axisymmetric conical form \citep{okazaki08,parkin09}. Based on the expected location of the cone apex at a certain orbital  phase and  following the wind-wind momentum balance \citep{canto96,okazaki08}, we place the apex of the cavity at a distance $d_{\mathrm{apex}}$ from the primary star. The density structure is axisymmetric about the $x$ axis. Our implementation is flexible enough to use different parameterizations of the density structure as input.  We computed limited 2-D models using the \citet{canto96} analytic solution as input for the geometry of the cavity, which yielded similar results as those discussed in this paper.

The 2-D models cover the same spatial scale as the 1-D CMFGEN model, extending up to 5500~AU from \etaa. Note that the long-term, large-scale effects caused by orbital motion predicted by 3-D hydrodynamical simulations \citep{madura11} are not included in our models. To properly solve the radiative transfer equation and still be feasible to compute a grid of models, we use typically 70 radial depth points and 51 angular points. Tests with up to 200 depth points and 90 angular points yielded similar results. 

The 2-D code of \citetalias{bh05} allows the emissivities and opacities to be either scaled or interpolated according to the modified density.  Both assumptions produce similar line profiles when including only the WWC cavity in the wind of Eta Car, and the precise choice does not affect the conclusions of this paper. Differences are only noticeable for the dense material in the WWC zone, in particular for $f_\alpha \ga 4$, when the interpolated source function predicts higher emission from optically-thick lines such as H$\alpha$. Hereafter, as a lower limit to the effects of the WWC zone, we present results for models with the scaled emissivities and opacities, except for the UV resonance and forbidden lines, where the interpolation of the source function is a more appropriate approximation. The observer's frame spectrum is computed using {\sc CMF\_FLUX}  \citepalias{bh05}.

The geometry of the orbit is assumed as in \citet{madura11}, with \etab\ located in the $x-y$ plane defined by the semi-major axis $x$ and semi-minor axis $y$, and the orbital axis $z$ being perpendicular to both. At apastron, \etab\ is located to the $+x$ direction. The observer is located at an inclination angle $i$, which is the tilt angle between the orbital axis and the line-of-sight ($i = 0^{\circ}$: face-on view, $i = 90^{\circ}$: edge-on view), and  views the binary system according to a longitude of periastron $\omega$ and at a certain orbital phase $\phiorb$. Periastron passage of the companion is defined as $\phiorb=0.0$ and apastron as $\phiorb=0.5$. For simplicity, hereafter we assume that phase zero of the spectroscopic cycle coincides with periastron passage. Note that the precise timing of periastron passage has not yet been constrained, but this does not affect our analysis of apastron observations.

Our fiducial 2-D cavity model of Eta Car assumes the stellar and wind parameters of $\etaa$ discussed in Sect. \ref{spherical}, while we adopt the orbital parameters and cavity orientation  from \citet{madura11} and references therein. We also assume the half-opening angle of the WWC cavity correspondent to a momentum ratio of $\eta=0.18$ derived from the X-ray analysis of \citet{parkin11}. Table \ref{params} presents the parameters of our fiducial 2-D model that, unless noted otherwise, are assumed in the next Sections.

\begin{table}
\caption{Parameters of our fiducial 2-D cavity model of Eta Car}
\label{params}
\begin{center}
\begin{tabular}{l c c}\hline
  \etaa's Stellar and Wind parameters & Value & Ref. \\ \hline
  Luminosity \lstar\ (\lsun)		& $5 \times 10^6$  & DH97\\
  Temperature \tstar\ (K, at $\tauross=130$)                 & 35200 & this work \\
  Temperature \teff\ (K, at $\tauross=2/3$)                    & 9400 & this work \\
  Radius \rstar\ (\rsun, at $\tauross=130$)                 & 60 & H01 \\
  Wind Terminal Velocity (km s$^{-1}$) & 420 & this work \\
  Mass-Loss Rate ($\times10^{-4}~M_{\odot}$ yr$^{-1}$) & $8.5$ & this work \\
  Volume filling factor $f$                     & 0.1 & H01\\
  Distance to Eta Car $d$	(kpc)				& 2.3 & AH93, S06\\
  \hline Cavity and orientation parameters  & Value & Ref.\\
  \hline
  Half-opening angle of the cavity $\alpha$ & 57\degr & M12,P11\\
  Width of the cavity walls $\delta_\alpha$ & 7\degr & this work\\  
  Overdensity of the cavity walls $f_\alpha$ & 4 & this work\\
  Density ratio within the cavity $b$ & 0.0016 & this work\\
  Distance $d_{\mathrm{apex}}$ at apastron & $23~\mathrm{AU}$ & O08, M12\\
  Orbital Inclination $i$ & $138^{\circ}$ & M12\\
  Longitude of Periastron $\omega$ & $ 260^{\circ}$ & M12\\
  P.A. on Sky of Orbital Axis $\mathrm{PA}_{z}$ & $\ 312^{\circ}$ 
  & M12\\\hline
\end{tabular}
\end{center}
References: DH97=\citet{dh97}, H01=\citet{hillier01}, O08=\citet{okazaki08}, M12=\citet{madura11}, P11=\citep{parkin11}.
\end{table}

\section{Effects of the WWC cavity on spectral lines of \etaa} \label{standardparamsline}

\begin{figure*}
\resizebox{\hsize}{!}{\includegraphics{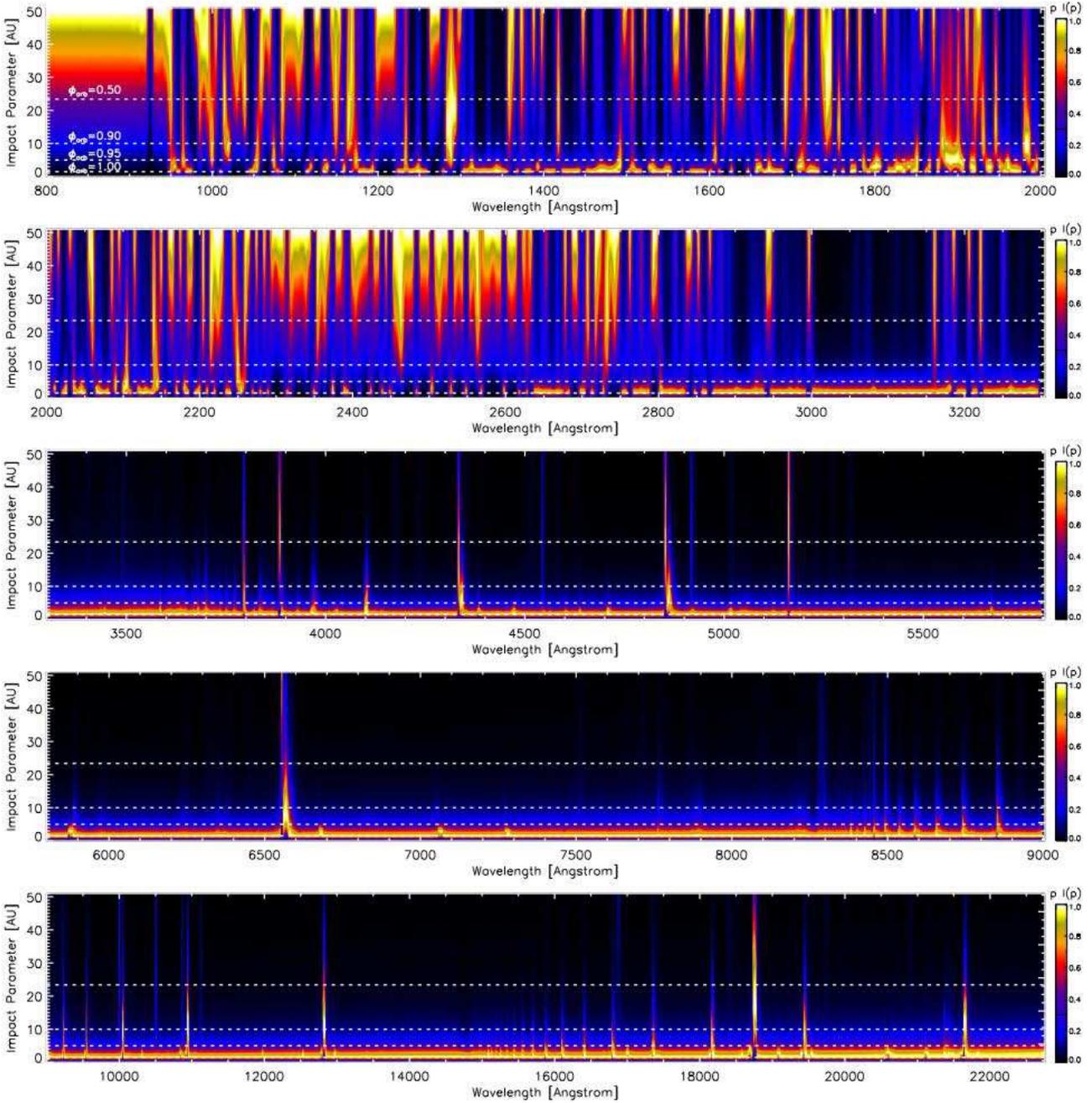}}
\caption{\label{pipimages} Variation of the flux-like quantity $p I(p)$ (color coded, normalized at each wavelength, and smoothed for clarity) as a function of impact parameter and wavelength for the 1-D CMFGEN model of \etaa. The horizontal dashed lines in all panels correspond to the distance of the apex of the WWC cavity to $\etaa$ for (from top to bottom) $\phiorb=0.5$, 0.9, 0.95, and 1.0. Spectral lines that have a significant fraction of $p I(p)$ originating above the horizontal dashed line will likely be affected at that \phiorb.}
\end{figure*}

Because the wind of Eta Car is extremely dense, the continuum and line formation regions are much more extended than the size of the hydrostatic core \citepalias{hillier01,hillier06}. In addition, the formation region of spectral lines is more extended than that of the continuum in the UV and optical. At these wavelengths, we anticipate that spectral lines will be more affected by the WWC cavity at apastron than the continuum. Here, we investigate thoroughly the reasons behind these effects since, in stellar winds, the formation of spectral lines is significantly more complex than the continuum formation. 

The formation region of spectral lines depends not only on the line opacity and emissivity, but on the energy levels involved in the transition, ionization potential (IP) of the species, non-LTE effects, and the line formation mechanism (i.e, recombination, resonance, continuum/line fluorescence, etc; see e.g., Mihalas 1978). In dense winds such as that from \etaa, the ionization structure is stratified, and generally lines from species with high IP form closer to the star than lines from elements with low IP \citepalias{hillier01}. Therefore, spectral lines that are formed over a larger region of the wind of \etaa\ are expected to be more affected by the presence of the WWC cavity \citepalias{hillier01}. Spectral lines formed in the innermost regions of \etaa's wind, where the WWC cavity does not penetrate, will not be greatly affected.

As an initial assessment of which spectral lines will be affected by the WWC cavity, we investigate the spatial region where the flux from different spectral lines originate in the 1-D CMFGEN model of \etaa. Recalling that, in a spherically symmetric outflow, the flux in the observer's frame at a given frequency $\nu$ is (e.g., Mihalas 1978)
\begin{equation}
F_{\nu}=\frac{2\pi}{d^2}\int^\infty_0{p\,I_\nu(p)\,dp},
\end{equation}
we investigate the behavior of the flux-like quantity $p\,I_\nu(p)$ as a function of impact parameter $p$, from the far-UV to the near-IR (Fig.~\ref{pipimages}). This provides a first estimate of which lines will be affected by the cavity, since {\it lines which have a significant fraction of their flux coming from $p>d_{apex}$ will be the ones more affected by the cavity.}    

Figure~\ref{pipimages} shows that the WWC cavity should affect the UV region even during apastron, especially in regions dominated by \ion{Fe}{ii} lines (e.g, $\lambda\lambda1000-1250$, $\lambda\lambda1350-1390$, $\lambda\lambda1410-1460$,  $\lambda\lambda2000-2600$). This happens because, for \etaa, \ion{Fe}{ii} lines are formed over very large spatial scales  \citepalias{hillier01,hillier06}. In the optical and near-IR, the main lines that have formation regions larger than $d_{apex}$ at apastron are H$\alpha$, H$\beta$, and  \ion{Fe}{ii} lines. 

However, we stress that, because $I(p)$ is a quantity projected on the sky, the actual magnitude of the effects  depends on the orientation of the WWC cavity. In addition, because the spectral lines are formed in an expanding outflow, different velocity regions of a spectral line form at different spatial regions (Fig. \ref{pipimages}). We find that spherical symmetry is broken by the presence of a WWC cavity and, depending on its orientation, different velocity regions of a spectral line are affected differently by the presence of low-density material within the cavity. {\it As a consequence, the presence of the WWC cavity causes certain spectral lines to be latitude and azimuth dependent, in particular H$\alpha$, H$\beta$, and  \ion{Fe}{ii} lines.}

\begin{figure}
\center
\resizebox{0.8\hsize}{!}{\includegraphics{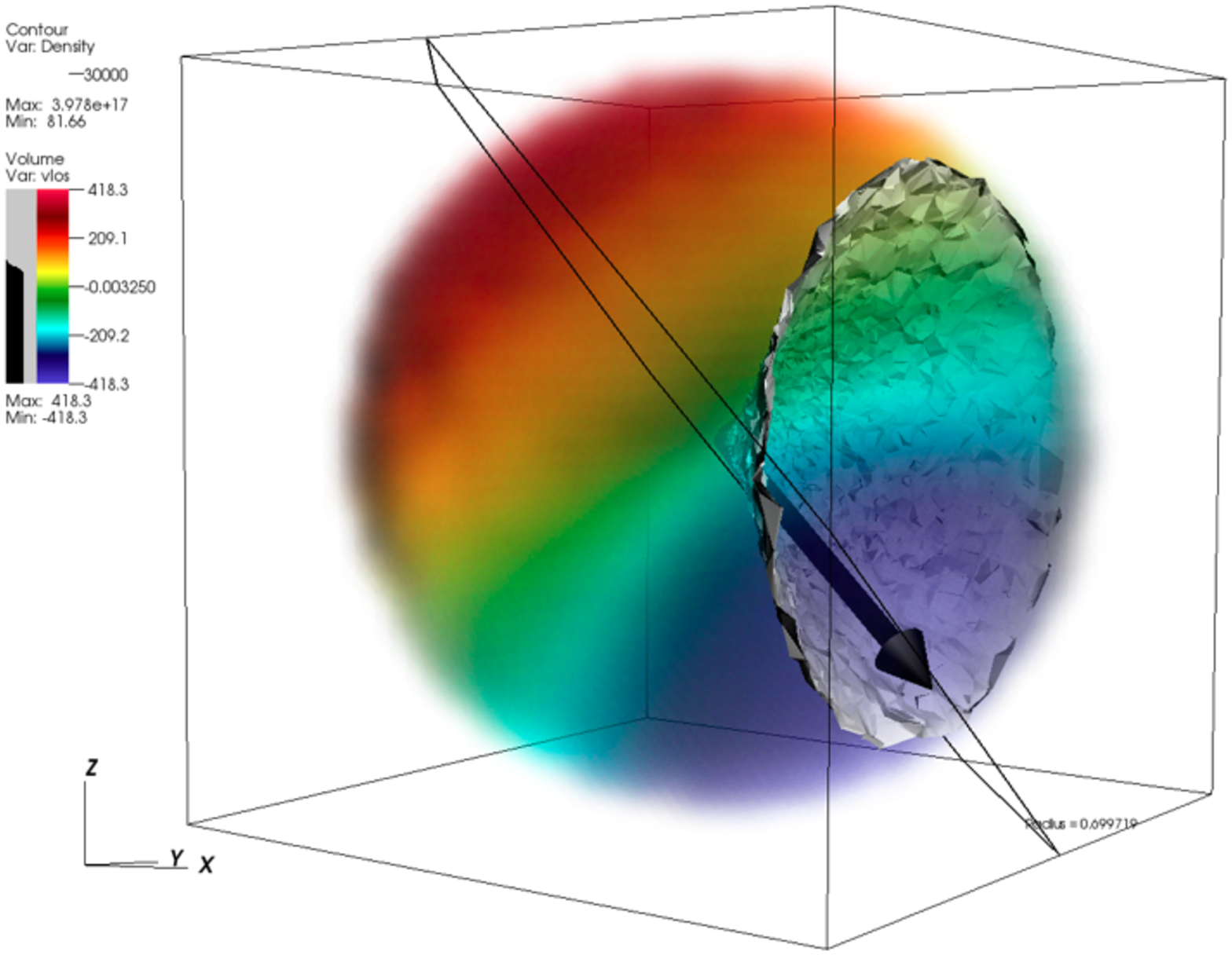}}
\resizebox{0.8\hsize}{!}{\includegraphics{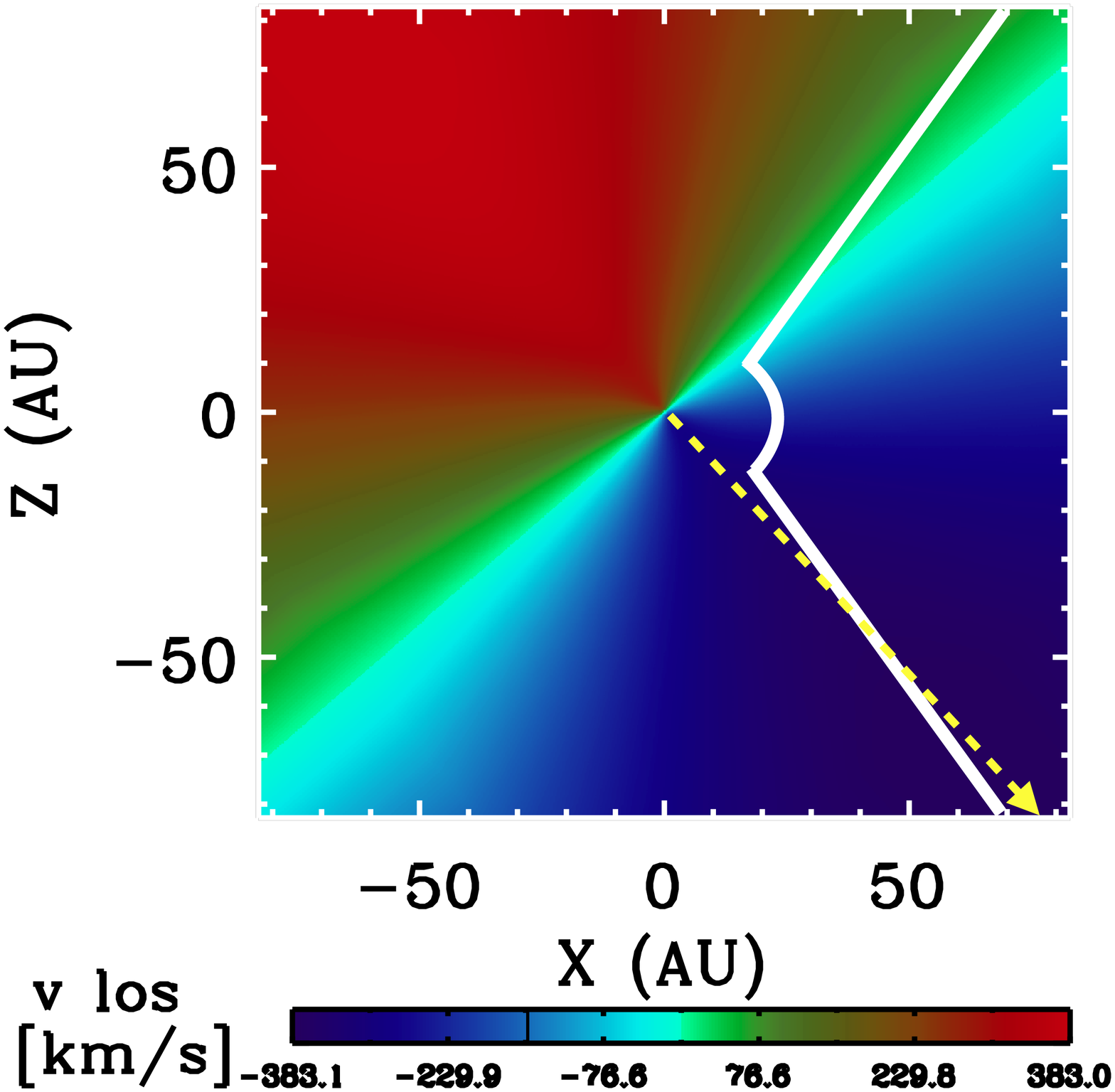}}
\caption{\label{vellos} Line-of-sight velocity of the 1-D spherical CMFGEN model of \etaa\ for an orbital orientation with $i=138\degr$ and $\omega=260\degr$. {\it Top:} Three-dimensional rendering, with the thick black arrow pointing toward the observer. The WWC cavity is shown in shades of gray. Note that this 3-D visualization does not correspond to the orientation on the sky, but it is rotated to better show the velocity regions of the wind of \etaa\ that are affected by the WWC cavity. {\it Bottom:} Cut in the x-z plane, with the yellow dashed arrow showing the direction to the observer, and the white solid line depicting the shape of the WWC cavity.}
\end{figure}

Moreover, as \etab\ moves in its eccentric orbit, $d_{apex}$ and the orientation of the WWC cavity change. Therefore, spectral lines that are not expected to be affected by the WWC cavity at apastron, such as the higher Balmer and Paschen lines, could be affected at phases around periastron. This occurs because higher Balmer and Paschen lines form in the inner, undisturbed regions of the wind of $\etaa$ around apastron, where the WWC cavity does not penetrate. At later $\phiorb$, however, $d_{apex}$ is smaller and the WWC cavity likely disturbs the formation region of the higher Balmer and Paschen lines. In addition, because of the change in the WWC cavity orientation as a function of $\phiorb$, lines that were affected at apastron at a given velocity interval will be affected at different velocity intervals around periastron. Ultimately, the presence of the WWC cavity will produce {\it phase-locked variations in spectral lines}.

Here we focus on the effects of the WWC cavity at phases around apastron. For a WWC cavity orientation with $i=138\degr$ and $\omega=260\degr$, our 2-D models show that the wind of \etab\ carves mainly the front-side of the wind of \etaa, which contains material outflowing toward the observer (Fig.~\ref{vellos}). Only a relatively minor portion of \etaa's wind flowing tangentially to or away from the observer is affected by the WWC cavity. {\it Therefore, the blue-shifted part of the line profiles should be the most affected by the WWC cavity during apastron.}

Indeed, this scenario is confirmed by our 2-D radiative transfer calculations. For orbital orientations corresponding to the observer looking down the rarified cavity at apastron, the  amount of Fe$^+$ in the line-of-sight is extremely reduced, causing less absorption of continuum radiation by \ion{Fe}{ii} transitions. Therefore, the 2-D cavity model has much less iron blanketing than the 1-D CMFGEN model.This effect is most pronounced in spectral regions full of  \ion{Fe}{ii} lines, such as around 1350--1450 ~\AA\ and 1800--2600~{\AA}. The extreme line overlap makes it complex to interpret individual features, which are usually blends of many spectral lines. 

\begin{figure*}
\resizebox{0.9\hsize}{!}{\includegraphics{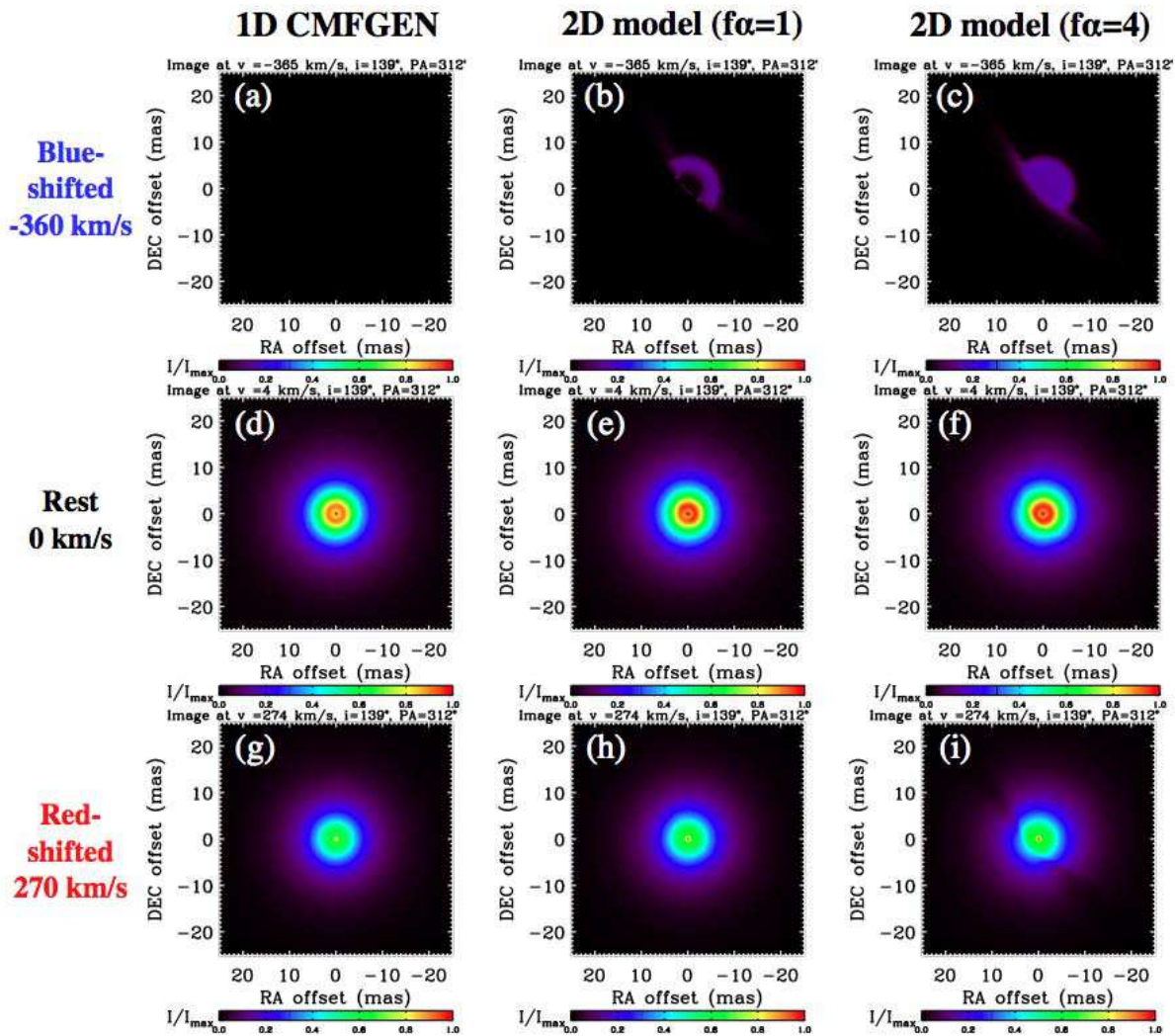}}
\caption{\label{iphalpha}{Monochromatic images of the H$\alpha$ line predicted by the 1-D CMFGEN model (left column), 2-D cavity model with the parameters from Table \ref{params} but without a dense WWC zone (i.e. $f_\alpha=1$, middle), and the 2-D cavity model with the parameters from Table \ref{params}. The different rows correspond to representative images in the blueshifted (top row, $v=-365~\kms$), zero velocity (middle row, $v=0~\kms$), and redshifted (bottom row, $v=274~\kms$) parts of the H$\alpha$ line profile. }}
\end{figure*}

Figure \ref{iphalpha} presents monochromatic H$\alpha$ images predicted by the 1-D and 2-D models at different velocities within the line. We investigate in detail the effects of the cavity for H$\alpha$, which is a recombination line formed over a large volume of the stellar wind. Analog effects are seen in the other unblended spectral lines, such as H$\beta$ and \ion{Fe}{ii} lines. 

We find that the 1-D model has strong blueshifted absorption at $p$ smaller than the photospheric size at 6564~{\AA} and no emission from larger $p$ (Fig. \ref{iphalpha}a). In contrast, the 2-D cavity model shows blueshifted emission at large $p$ in directions toward where the cavity is projected on the sky (NW in this case, Fig. \ref{iphalpha}c). A 2-D model including only the WWC cavity but no dense WWC zone  (Fig. \ref{iphalpha}b) presents a similar behavior, showing that the key factor is the presence of a WWC cavity in the wind of \etaa. The enhanced blueshifted emission at large $p$ is caused by the reduced optical depth because of the presence of the WWC cavity, allowing line photons that would otherwise be absorbed by the wind of \etaa\ to escape.  Since the flux at a give wavelength is $F=\int{}{I(p) dp d\delta}$, the net result is a significant decrease in the amount of blueshifted H$\alpha$ absorption. 

The WWC cavity also has a moderate effect in the emission at zero velocities in H$\alpha$, allowing more line photons to escape toward the observer, but does not significantly change the image morphology (Figs.  \ref{iphalpha}e,f).  Because the back side of the primary wind is essentially not carved by the wind of \etab\ around apastron and remains undisturbed, no effects are seen in the redshifted emission of  H$\alpha$ when considering the WWC cavity alone (Fig. \ref{iphalpha}h).

In addition to the WWC cavity, the line emission is also affected by the  presence of  a dense, shocked WWC zone. We find that, because of the orientation of the cavity, blueshifted H$\alpha$ emission from the WWC zone is roughly aligned with the NE-SW direction, while redshifted emission is higher to the NW (Fig. \ref{iphalpha}f,i). This morphology resembles what has been seen in the extended forbidden line emission that originates in the outer, time-averaged WWC zone (Gull et al. 2011), but scaled down by a factor of 30 spatially. 

Our models predict that H$\alpha$ is optically thick, with a Sobolev optical depth larger than 1000 in the inner wind (for $r \la 200$ AU). Because of that, the zero and redshifted images of  H$\alpha$ are also affected by absorption of line photons from the {\it back side} of the primary wind by the WWC zone. This effect is more pronounced for large $f_\alpha$. The absorption occurs mainly along the NE-SW direction (Fig. \ref{iphalpha}h,i), and is caused by the redshifted line-of-sight velocity of material in the WWC zone toward these directions. 

\section{Comparison with observations around apastron}\label{compareobs}

Building on the insights from previous Sections, we compare below the fiducial 2-D cavity model spectrum to the \hststis\ observations obtained at $\phisp=10.410$ and 10.603. As discussed below, our 2-D models show the presence of a rarified WWC cavity in the dense wind of \etaa\ is sufficient for reproducing the UV and optical spectra of the central source, without the need of a rapid-rotating \etaa. At this point, however, we cannot exclude that both scenarios occur simultaneously, with \etaa\ being a rapid rotator.

\begin{figure*}
\resizebox{\hsize}{!}{\includegraphics{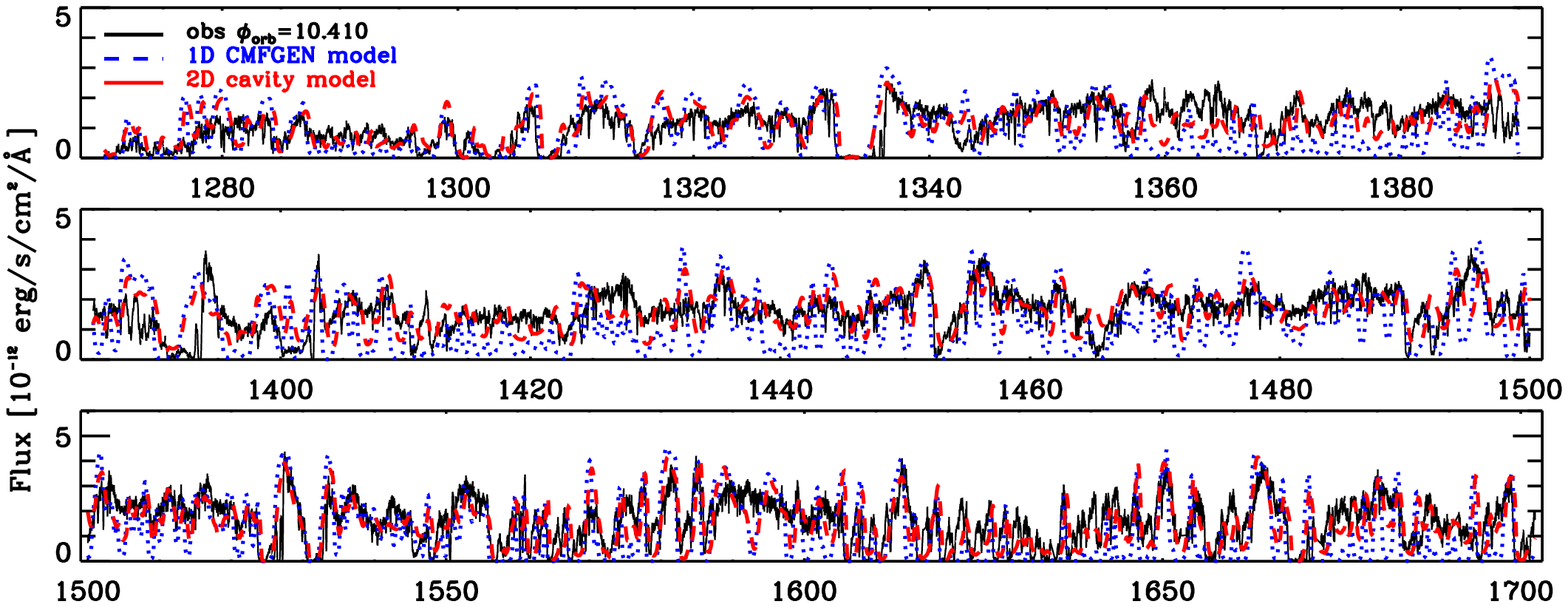}}
\resizebox{\hsize}{!}{\includegraphics{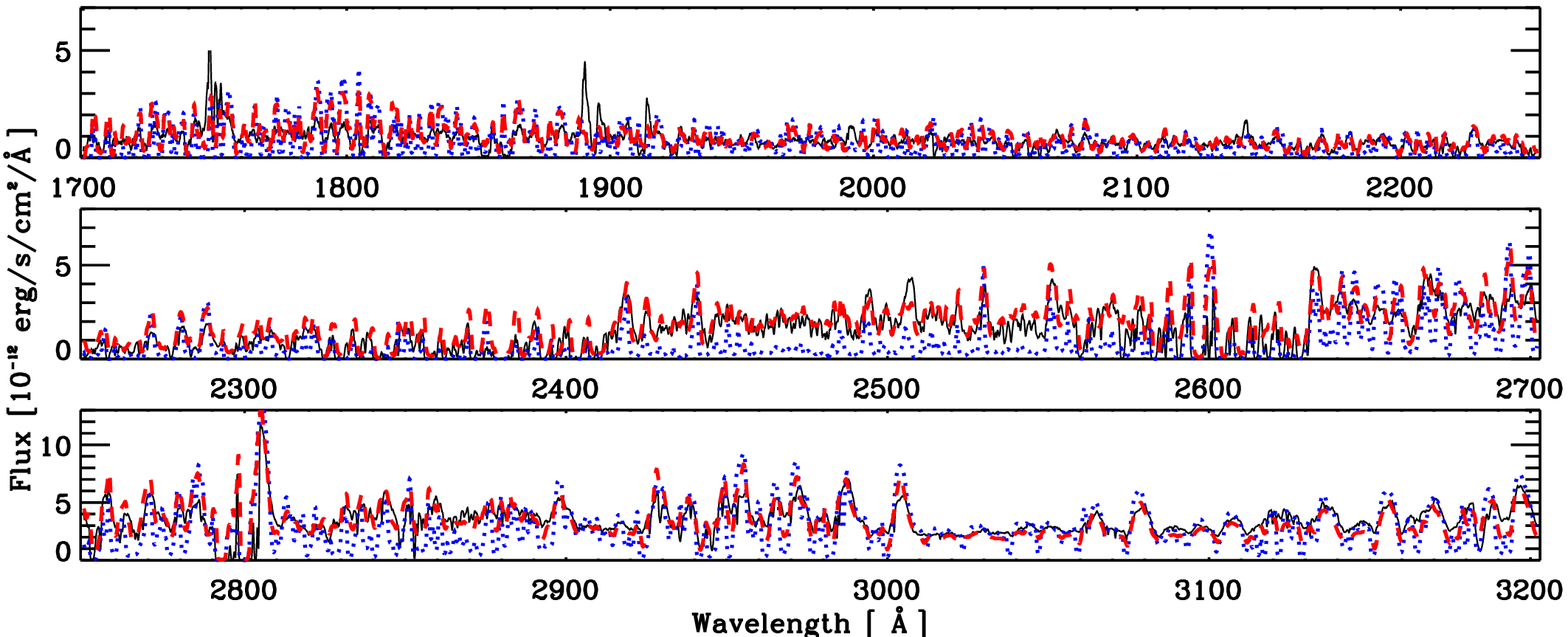}}
\caption{\label{moduvstandard} Comparison between the UV spectrum of Eta Car in the range $1260-3200$~\AA~observed with \hststis\ at $\phisp=10.410$ (solid black line), the 2-D model including the cavity due to the wind-wind interaction (red dashed line), and the 1-D CMFGEN model of \etaa\ without the cavity (dotted blue line). Both models were reddened according to the interstellar extinction law of \citet{fitzpatrick99} with $R_\mathrm{V}=4.0$ and $E\mathrm{(B-V)}=1.0$, and scaled to match the observed continuum \citepalias{hillier01}. }
\end{figure*}

\subsection{Ultraviolet spectrum}

The UV spectrum of Eta Car is strongly dominated by bound-bound transitions of \ion{Fe}{ii} believed to originate in the wind of \etaa\ \citepalias{hillier01,hillier06}.  While both the revised spherical model presented here and the one from \citetalias{hillier01} have difficulties fitting the UV spectrum of Eta Car (unless light from the inner $\sim0\farcs033$ of the model is blocked; \citetalias{hillier06}), our 2-D model reproduces reasonably well the UV spectrum (Fig.~\ref{moduvstandard}). The quality of the fit improves compared to those from the spherical model because, for the orbital parameters assumed here (Table \ref{params}), the observer looks down the low-density cavity, which dramatically reduces the column density, and thus opacity and optical depth of \ion{Fe}{ii} transitions in the line-of-sight to \etaa. The end result is less absorption of  \ion{Fe}{ii} in the UV spectrum {\it while} still maintaining the $\mdot$ of \etaa\ at $8.5\times10^{-4}~\msunyr$. Therefore, according to our models, the UV spectrum is satisfactorily explained by the carving of the wind of \etaa\ by \etab, without the need of  a ``coronagraph" blocking the inner 0\farcs033, as proposed by \citetalias{hillier06}. Anomalous extinction could still occur in the line-of-sight to the central source (Sect. \ref{conc}). The fact that our 2-D models reproduce well the UV spectrum indicates that \etab\ has little flux contribution in the wavelength range analyzed here (1250-3300~\AA). This suggests that, perhaps, the luminosity of \etab\ is much lower than previously thought, maybe around a few times $10^5~\lsun$ and consistent with the lower range of values suggested by \citet{mehner10}. Further updates of our code to include the flux contribution of \etab\ are needed to improve knowledge on \etab.

Our model fails to reproduce high-ionization resonance lines, such as \ion{Si}{iv} $\lambda\lambda~1398,1402$ and \ion{C}{iv} $\lambda\lambda~1549,1551$ (Fig.~\ref{moduvstandard}). The P-Cygni absorption in these lines extends up to $\sim-750~\kms$ around apastron \citep{hillier06,gnd10}. These velocities are consistent with the high-ionization resonance lines being formed in a region dominated by the turbulent velocity field of \etaa's wind or in the WWC zone, perhaps moving slightly faster than the terminal speed of \etaa. As discussed in \citet{gnd10}, we speculate that these high-ionization resonance lines arise in modified regions of the wind of \etaa, close to the WWC zone, that are photoionized by \etab\ (see Sect. \ref{ionization}).

\subsection{Optical spectrum}

\begin{figure*}
\resizebox{\hsize}{!}{\includegraphics{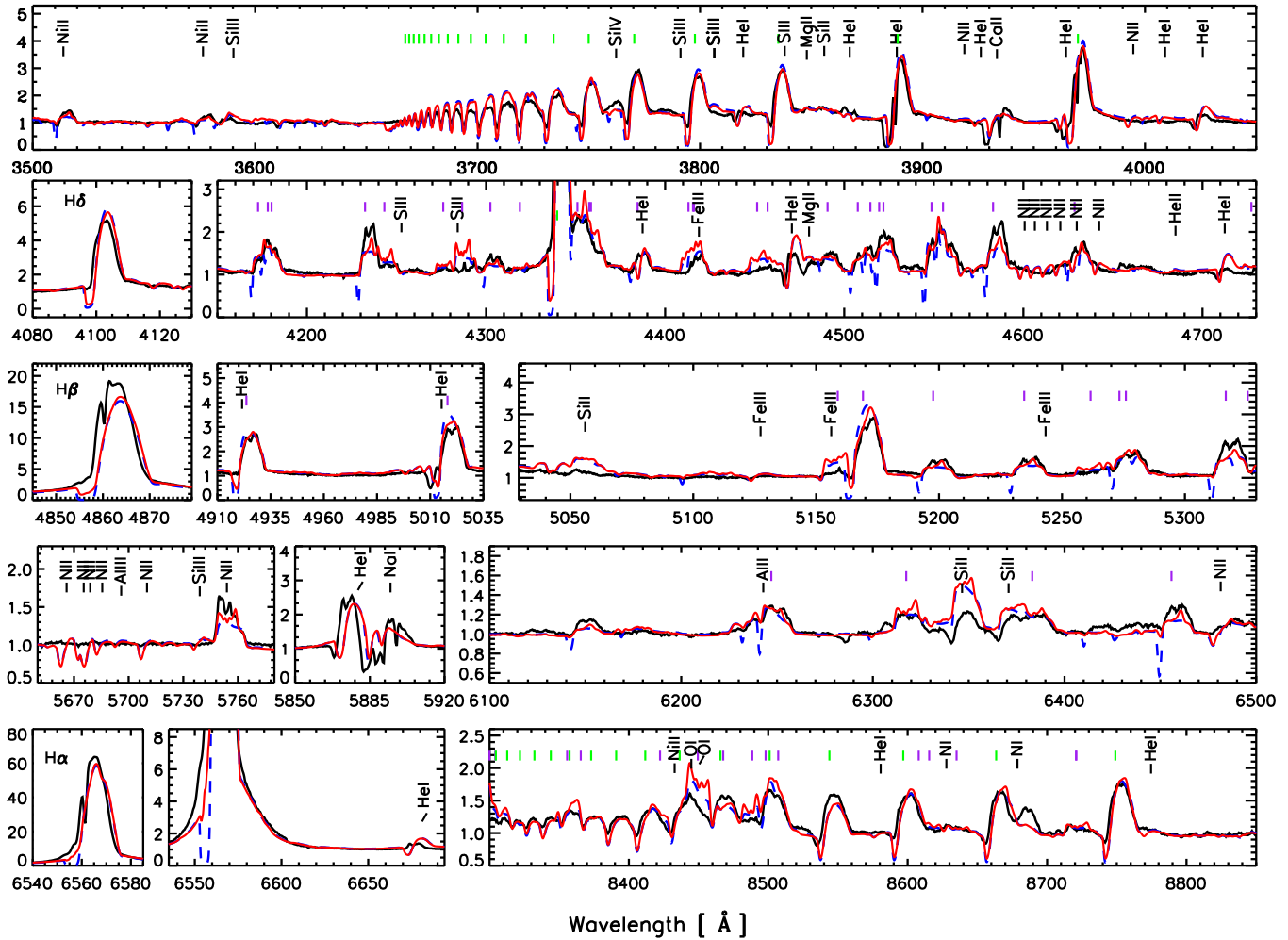}}
\caption{\label{modoptstandard} Comparison of the spectrum of Eta Car observed with \hststis\ in 2001 Apr 17 ($\phisp=10.603$;  black solid line) with the 2-D CMFGEN models including the cavity due to the wind-wind interaction (red solid line) and the 1-D CMFGEN model of \etaa\ without the cavity (dashed blue line). The fluxes are continuum-normalized. The main spectral lines are identified, with green and purple ticks marking the laboratory wavelength of  {\sc H} and \ion{Fe}{ii} transitions, respectively.}
\end{figure*}

The optical spectrum of the central $\sim0\farcs2$ of Eta Car is dominated by strong emission lines from {\sc H} and \ion{Fe}{ii} \citep{davidson95,hillier01} that were shown to arise in the wind of \etaa\ \citepalias{hillier01}. While the spherical model from \citetalias{hillier01} strongly overestimates the amount of P-Cygni absorption in these lines, our 2-D model yields much less P-Cygni absorption in H$\alpha$, H$\beta$, H$\gamma$, and \ion{Fe}{ii} lines (Fig.~\ref{modoptstandard}). The explanation for the better fits is similar to that of the UV spectrum: since the observer looks down the cavity, there is less  column density and optical depth in the line-of-sight. However, the 2-D synthetic spectrum still presents more absorption than seen in the observed spectrum, in particular in  H$\delta$ and higher Balmer and Paschen lines. The absorption part of these lines arises in regions of the primary wind unaffected by the WWC cavity at apastron, and thus 2-D and 1-D models have similar line profiles. A plausible explanation for the excess absorption is that the 2-D models currently neglect photoionization of the wind of \etaa\ by \etab. The assumed orbital and WWC cavity parameters could also be different (see Sects. \ref{orbstudy} and \ref{paramstudy}). 

Weak  \ion{N}{ii}  and  \ion{He}{i} lines are also present in the optical spectrum and most authors concur these are, at least during some orbital phases, affected by photoionization of \etab\ \citep{damineli97,davidson05,nielsen07,humphreys08,mehner11}.  In this case, the formation regions of these lines depend on the number of ionizing photons from \etab, which depends mainly on its temperature, luminosity, and mass-loss rate.  We found that  the inclusion of the rarified WWC cavity and a dense WWC zone in the model is insufficient to provide better fits for these lines. As for the higher Balmer lines, the \ion{He}{i} emission and absorption that originates in the primary wind, if any, is unchanged by the presence of the WWC cavity alone at apastron (Fig.~\ref{modoptstandard}). This is because the WWC cavity does not penetrate in the region where \ion{He}{i} lines are formed.

\section{Constraints on the orbital parameters} \label{orbstudy}

As shown in Sect. \ref{standardparamsline}, the UV and optical spectra of \etaa\ are severely affected by the presence of the rarified WWC cavity carved by \etab\ and the dense WWC zone. In this Section we analyze the constraints that can be imposed on the orbital inclination angle $i$ and the longitude of periastron $\omega$. For brevity, we quote results for selected spectral regions of \etaa's spectrum: the near-UV ($\lambda\lambda2400-2600$), H$\alpha$, H$\beta$, H$\delta$, and optical \ion{Fe}{ii} lines  ($\lambda\lambda4500-4600$). Note that because we analyze spatially unresolved spectra from the central source, our models are insensitive to the value of PA$_z$. Interferometric observations that spatially resolves  lines formed in the stellar wind of \etaa, such as Bracket $\gamma$ \citep{weigelt07}, could be used to further constrain PA$_z$.

 \subsection{The orbital inclination angle $i$}

\begin{figure}
\resizebox{\hsize}{!}{\includegraphics{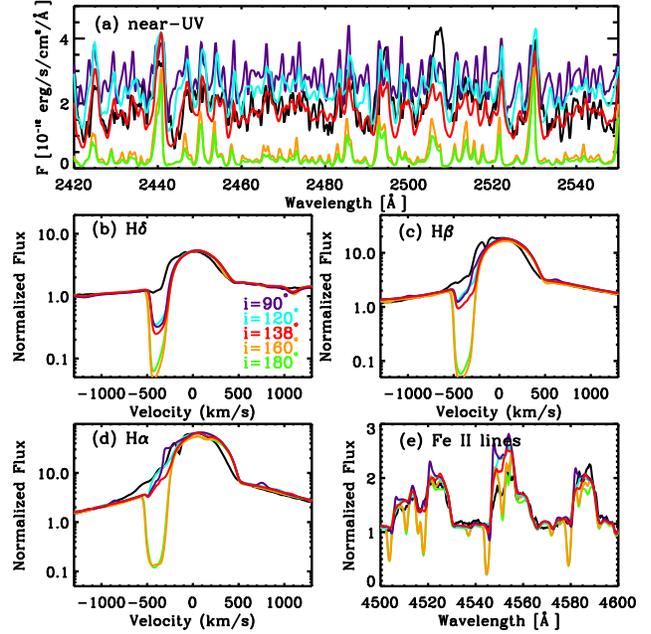}}
\caption{\label{figincwall}Spectra from 2-D cavity models of Eta Car computed for different $i$, assuming $\alpha=57\degr$ and $\omega=260\degr$. The panels show (a) the near-UV  region around $\lambda2500$, (b) H$\delta$, (c)  H$\beta$, (d) H$\alpha$, and (e) optical \ion{Fe}{ii} lines in the region $\lambda\lambda4500-4600$. Note the latitudinal dependence on all spectral lines. In all panels, the black solid line corresponds to the observed \hststis\ spectrum at $\phisp=10.603$.}
\end{figure}

Our 2-D models indicate that the UV spectrum and the morphology of spectral lines significantly change as a function of $i$. Figure~\ref{figincwall} presents a set of 2-D models computed for different $i$, with the other parameters kept fixed (Table \ref{params}). Orbital orientations with $i$ and $180\degr-i$ produce identical model spectra, with the only difference being a counter-clockwise or clockwise orbital motion projected on the sky, respectively. Because the spectra analyzed here are spatially unresolved, this degeneracy cannot be broken with our models. Therefore, given the evidence supporting a clockwise orbital motion on the sky \citep{smith04,gull09,madura11}, we investigate models with $90\degr \leq i \leq 180\degr$, noting that analogous constraints would be obtained for the parameter space $90\degr \geq i \geq 0\degr$.

As shown in Fig. \ref{figincwall} we find that most H and \ion{Fe}{ii} lines are latitude dependent. The strongest latitude-dependent effects are seen in regions of the UV spectrum dominated by \ion{Fe}{ii} lines and in H$\alpha$, H$\beta$, and optical \ion{Fe}{ii} lines. The differences seen in the UV spectrum and strength of the P-Cygni absorption of optical lines as a function of $i$ are both related to the varying amount of primary wind material in the line-of-sight, with the WWC cavity affecting different velocity regions of the line profiles (see Sect. \ref{standardparamsline}) depending on $i$. 

The theoretical profiles presented in Fig. \ref{figincwall} qualitatively resemble the observed line profiles scattered off different locations in the Homunculus nebula shown in \citet{smith03}.  {\it Our 2-D cavity models predict that spectra scattered off the Homunculus poles should have stronger P-Cygni absorption than those observed on the central star or equatorial positions. }The presence of the WWC cavity thus mimic observed effects that could, in principle, also arise in a single, fast-rotating \etaa. These have profound implications for interpreting the nature of the central source in Eta Car, and an in-depth investigation of this effect will be presented in an accompanying paper.

Here we focus on constraining $i$, using as diagnostics the amount of \ion{Fe}{ii} absorption in the UV and blueshifted absorption of H$\alpha$, H$\beta$, and \ion{Fe}{ii} lines. We find that orientations with $110\degr  \la i \la  140\degr $ seem to be required to fit the UV spectrum and H$\alpha$ line profiles satisfactorily  (Fig. \ref{figincwall}). These values of allowed $i$ correspond to orientations in which the observer looks down the low-density WWC cavity, with the amount of \ion{Fe}{ii} and H$\alpha$ P-Cygni absorptions reasonably matching the observations. 

The amount of primary wind material in the line-of-sight depends not only on $i$ but also on $\omega$ and $\alpha$. Variations in $\alpha$ could change the range of $i$ that would fit the UV and optical apastron spectra according to
\begin{equation}
 110\degr \la i \la  (90\degr + \alpha),
 \end{equation} 
for $25\degr \la \alpha \la 90\degr$.

In addition to affecting the absorption profiles, changing $i$ might cause the emission line profiles to present sub-structures which vary in intensity and velocity. These are caused by the emission and absorption of dense, shocked material along the WWC zone (see Sect. \ref{standardparamsline} and Fig. \ref{iphalpha}).

\subsection{Longitude of periastron $\omega$}

\begin{figure}
\resizebox{\hsize}{!}{\includegraphics{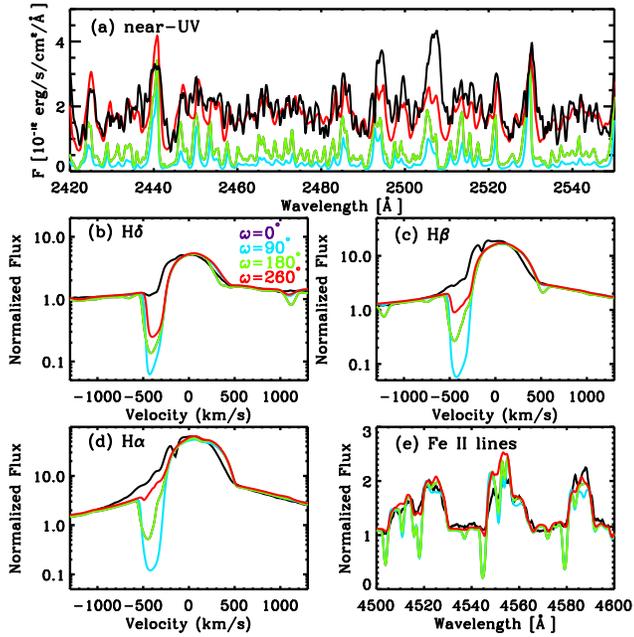}}
\caption{\label{figomega}Similar to Fig. \ref{figincwall}, but  for different $\omega$, assuming $\alpha=57\degr$ and $i=138\degr$. Note that the line profiles for $\omega=0\degr$ and $180\degr$ are very similar and are superimposed. }
\end{figure}

We find that the presence of a WWC cavity induces not only latitudinal variations in some of the line profiles of Eta Car, but also azimuthal variations that can be used to constrain the longitude of periastron $\omega$ (Fig. \ref{figomega}). Similar to the effects of $i$ and $\alpha$, the value of $\omega$ also regulates the amount of primary wind material in the line-of-sight, and the amount of absorption in H$\alpha$, H$\beta$, and \ion{Fe}{ii} lines, among others.

Our 2-D models indicate that the best fit to the UV and optical spectra at apastron occurs for orientations with $210\degr \la \omega \la 330\degr$, assuming $\alpha=57\degr$ and $i=138\degr$, corresponding to the observer viewing the binary system down the WWC cavity. These values are consistent with the 3-D orbital parameters found by \citet{madura11}. Changing $i$ and $\alpha$ could slightly affect the allowed ranges of $\omega$, but will not allow models with, e.g. $\omega=90\degr$ \citep{abraham05b,kashi08a}, to be consistent with the observed spectrum.

\section{Constraints on physical properties of the WWC cavity} \label{paramstudy}

In this Section we investigate how structural changes in the WWC cavity, such as the half-opening angle and density of the WWC zone, affect the emerging UV and optical spectrum. In addition to obtaining insights on the physical properties of the system, long-term changes in $\mdot$ and/or $\vinf$ of one or both stars might lead to changes in the structure of the WWC cavity, which in turn could affect the line profiles. For brevity, we present results for H$\alpha$, but similar effects are seen in other H and \ion{Fe}{ii} lines.

\subsection{The half-opening angle of the cavity $\alpha$}

\begin{figure}
\resizebox{\hsize}{!}{\includegraphics{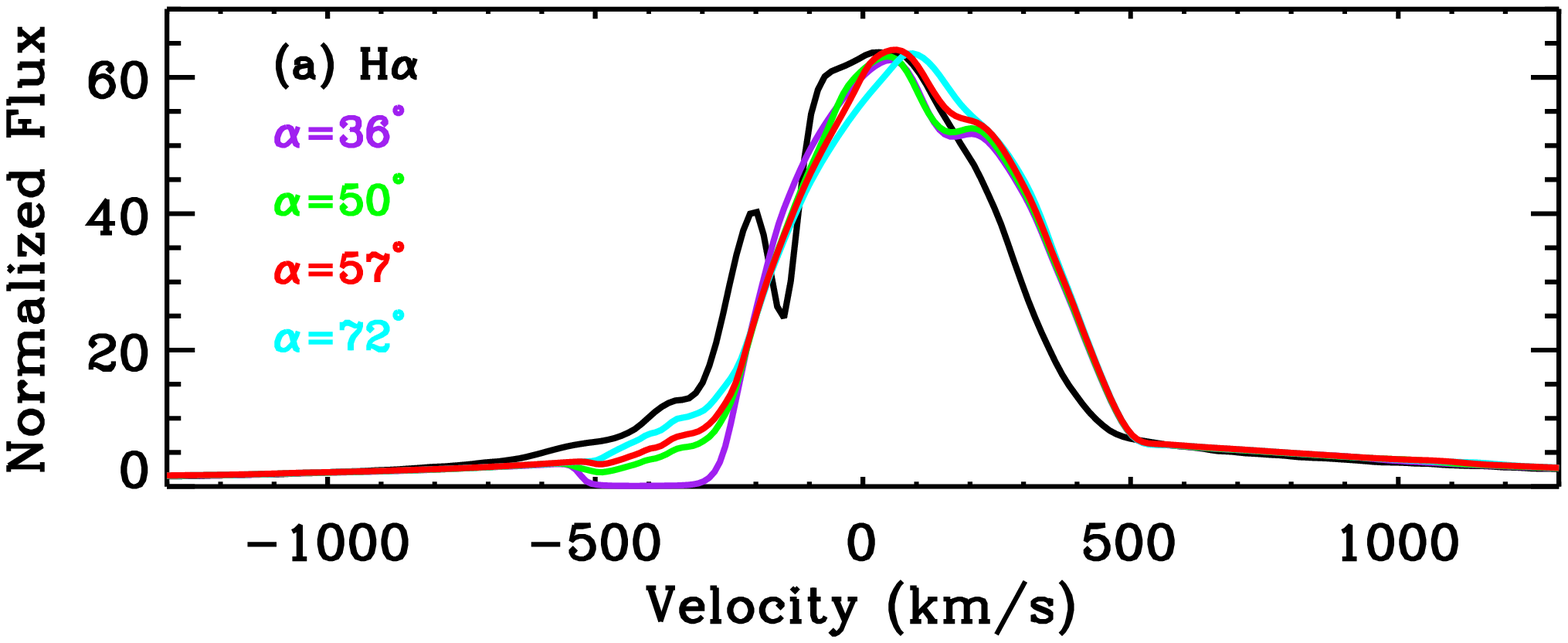}}
\resizebox{\hsize}{!}{\includegraphics{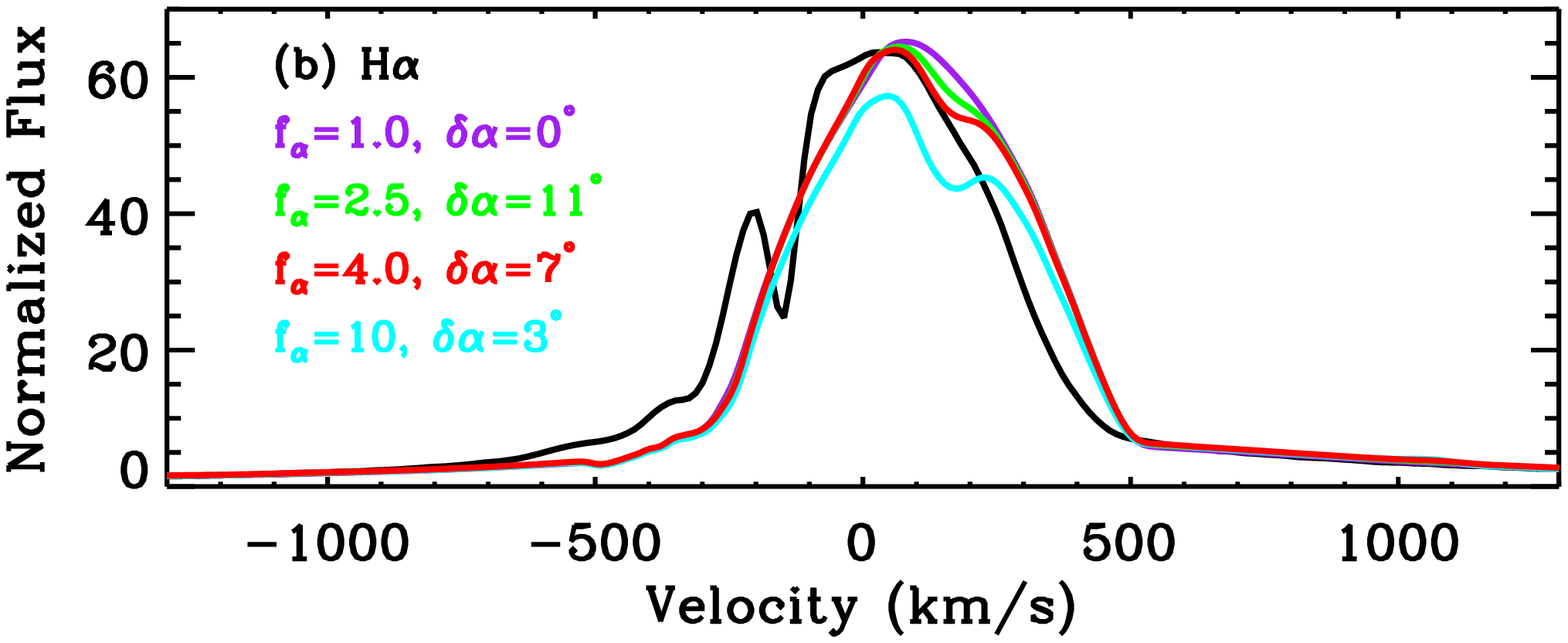}}
\caption{\label{figalpha}{\it(a):} H$\alpha$ line profiles computed with the 2-D cavity model assuming different $\alpha$, for $\omega=260\degr$ and $i=138\degr$. The observed \hststis\  spectrum at $\phisp=10.603$ is show in black.  {\it(b):} Similar to {\it(a)}, but for 2-D cavity models with different $f_\alpha$ and $\delta_\alpha$.}
\end{figure}

Figure~\ref{figalpha}a compares the \hststis\ observations at $\phisp=10.410$ with synthetic H$\alpha$ profiles from our 2-D models for different values of $\alpha$. The other model parameters are kept fixed (see Table \ref{params}).

We find that changing $\alpha$ has a dramatic influence in the absorption and low-velocity emission parts of the P-Cygni profile of H$\alpha$. Similar to $i$ and $\omega$,  $\alpha$ also affects the amount of wind of \etaa\ that is carved by \etab. The amount of carving determines whether the observer looks down the cavity or through the unmodified wind of \etaa, dramatically impacting the Hydrogen $n=2$ energy level population in the line-of-sight, which regulates the H$\alpha$ absorption. We found that 2-D models with $\alpha \la 50\degr$ show a strong, saturated P-Cygni absorption component, since the observer's line-of-sight crosses mainly primary wind material (Fig.~\ref{figalpha}a). In contrast,  for models with $\alpha \ga 50\degr$, the observer looks down the cavity and, consequently, H$\alpha$  shows weak or no P-Cygni absorption (Fig.~\ref{figalpha}a). Also, as before, a change in $i$ or $\omega$ would allow for slightly different $\alpha$. Further constraints on $\alpha$ will be obtained in future studies of the variation of spectral lines as a function of \phiorb.

The amount of H$\alpha$ emission changes only slightly for models with $\alpha=36\degr-72\degr$ (Fig. \ref{figalpha}a). This effect  occurs because the redshifted H$\alpha$ emission comes essentially from the back side of the wind of \etaa that,  for $i=138\degr$ and $\omega=260\degr$, is only weakly carved by the WWC cavity (Fig. \ref{vellos}). To see a significant effect in the emission strength, \etab\ would have to dominate the wind momentum balance in order to push the cavity to the back side of \etaa's wind (i.e., $ \alpha > 90\degr$).

Our 2-D model favors $50\degr \la \alpha \la 70\degr$ to fit the observations since, as mentioned before, the observed H$\alpha$ profile shows no P-Cygni absorption profile near apastron. A similar range of $\alpha$ is required to fit other optical lines and the UV spectrum simultaneously for $\omega=260\degr$ and $i=138\degr$. These values of $\alpha$ are consistent with those determined from analyses of the X-ray lightcurve \citep{pc02,okazaki08,parkin09}. However, the value of $\mdot$ of \etaa\  obtained here based on the UV and optical spectra ($8.5\times10^{-4}~M_{\odot}$ yr$^{-1}$) is higher than that derived based on X-rays ($4.8\times10^{-4}~M_{\odot}$ yr$^{-1}$, \citealt{parkin11}). Consequently, our derived value of $\alpha$ requires \etab\ to have  $\mdot\simeq2.0\times10^{-5}~\msunyr$ in order to achieve similar momentum ratios as deduced from X-ray observations.

In addition, the  value of $\alpha$ could be biased due to neglecting the photoionization of the wind of \etaa\ by \etab. In this scenario, portions of the wind of \etaa\ which would have neutral H in the absence of \etab\ are actually photoionized, reducing the population of the $n=2$ energy level of H and, consequently, the optical depth of H$\alpha$. This scenario might have important consequences, in particular if the observer's line of sight crosses a significant amount of the primary wind close to the interaction region, which would be the case for Eta Car if indeed $i\simeq138\degr$ and $\alpha\simeq57\degr$.

\subsection{Density of the WWC zone $f_\alpha$}

Besides including microclumping via a volume filling factor $f$, our 2-D models assume a WWC zone that is smooth on large scales (Fig. \ref{cavity_sketch}). Therefore, the calculations do not include large scale, optically-thick clumps, as would be expected if the primary wind side of the shock cools radiatively \citep{pc02, parkin09,parkin11,moffat09}. In addition, the wind of $\etaa$ itself may contain large scale clumps, as suggested by polarimetric studies of LBVs \citep{davies05,davies07}. Having that caveat in mind,  we attempt here an exploratory investigation of the spectral line morphology for different densities and widths of the primary side of the WWC zone. We assume that mass conservation ties $f_\alpha$, $\delta_\alpha$, and $\alpha$ according to Equation \ref{eqfalpha}. 

Figure \ref{figalpha}b presents H$\alpha$ line profiles illustrating the effects of $f_\alpha$ and $\delta_\alpha$ on the spectrum of \etaa. We note that similar effects are seen in other optically-thick lines affected by the WWC cavity, such as H$\beta$ and strong \ion{Fe}{ii} lines. Our 2-D models show that most of the effects of varying $f_\alpha$ and $\delta_\alpha$ occur at low velocities, mostly on the redshifted side of the line profile. As  $f_\alpha$ increases, low-velocity redshifted absorption from the back wall of the WWC zone affects more and more the H$\alpha$ profile. Blueshifted emission and absorption from the front wall of the WWC zone may also affect the H$\alpha$ line profile. We find that models with $f_\alpha\simeq 1$ to 4 fit reasonably the H$\alpha$ line profile at apastron, while model spectra with $f_\alpha\ga 5$ present sub-structures in the line profile that are not seen in the observed spectrum.

\section{Effects of the WWC cavity on the continuum of \etaa} \label{standardparamscont}

In this Section, we analyze how the presence of the WWC cavity affects the continuum of \etaa\ during apastron. As we discuss below, these changes could occur due to the `bore-hole' phenomenon (Owocki \& Smith, priv. comm.; \citealt{madurathesis10}) and modification of the optical depth of \etaa's wind, and emission/absorption from the WWC zone \citep{gmo10}.

As found by \citetalias{hillier01} and \citetalias{hillier06}, the continuum extension of \etaa\ is regulated by the wavelength- and radially-dependent opacity and emissivity. \citet{madura10}, \citet{madurathesis10}, \citet{madura11a}, and Madura et al. (2012, in prep.) showed that, depending on the opacity and emissivity of the wind of \etaa, continuum flux changes at a given wavelength could occur when the apex of the WWC cavity reaches the surface of optical depth $\tau=2/3$ at this wavelength. This `bore-hole' phenomenon \citep{madura10} is likely related to the variations seen in optical and near-IR photometry around periastron \citep[e.\,g.,][]{whitelock04,lajus10}. We refer the reader to Madura et al. (2012, in prep.), where the continuum flux variation as a function of orbital phase is investigated in detail.

\begin{figure}
\resizebox{\hsize}{!}{\includegraphics{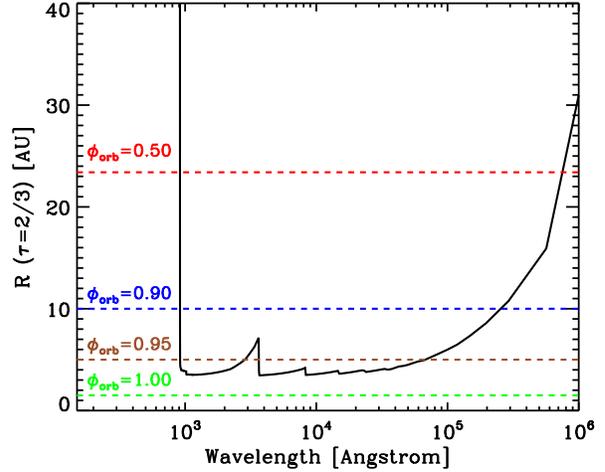}}\\
\caption{\label{taulambda}{Distance from \etaa\ where $\tau=2/3$ in the continuum as a function of wavelength (black solid line). The horizontal dashed lines correspond to the distance of the apex of the WWC cavity to $\etaa$ for $\phiorb=0.5$ (red), 0.9 (blue), 0.95 (brown), and 1.0 (green).}}
\end{figure}

Based on the revised 1-D CMFGEN model presented in Sect. \ref{lowervinf}, we investigate which wavelength regions could be affected by the cavity. Figure \ref{taulambda} presents the distance from \etaa\ where $\tau=2/3$ as a function of wavelength for our revised spherical model. We find that the apex of the WWC cavity does not reach optically-thick regions of \etaa's wind at most wavelengths for most of the orbital cycle.  Figure \ref{taulambda} shows that  at apastron the cavity reaches regions with $\tau\geq2/3$ only for $\lambda < 912~\ang$ or $\lambda > 80~{\micron}$. Only at $0.95 \la \phiorb \la 1.05$ does the cavity penetrate optically-thick regions in the optical and near infrared continuum. 

\begin{figure}
\resizebox{\hsize}{!}{\includegraphics{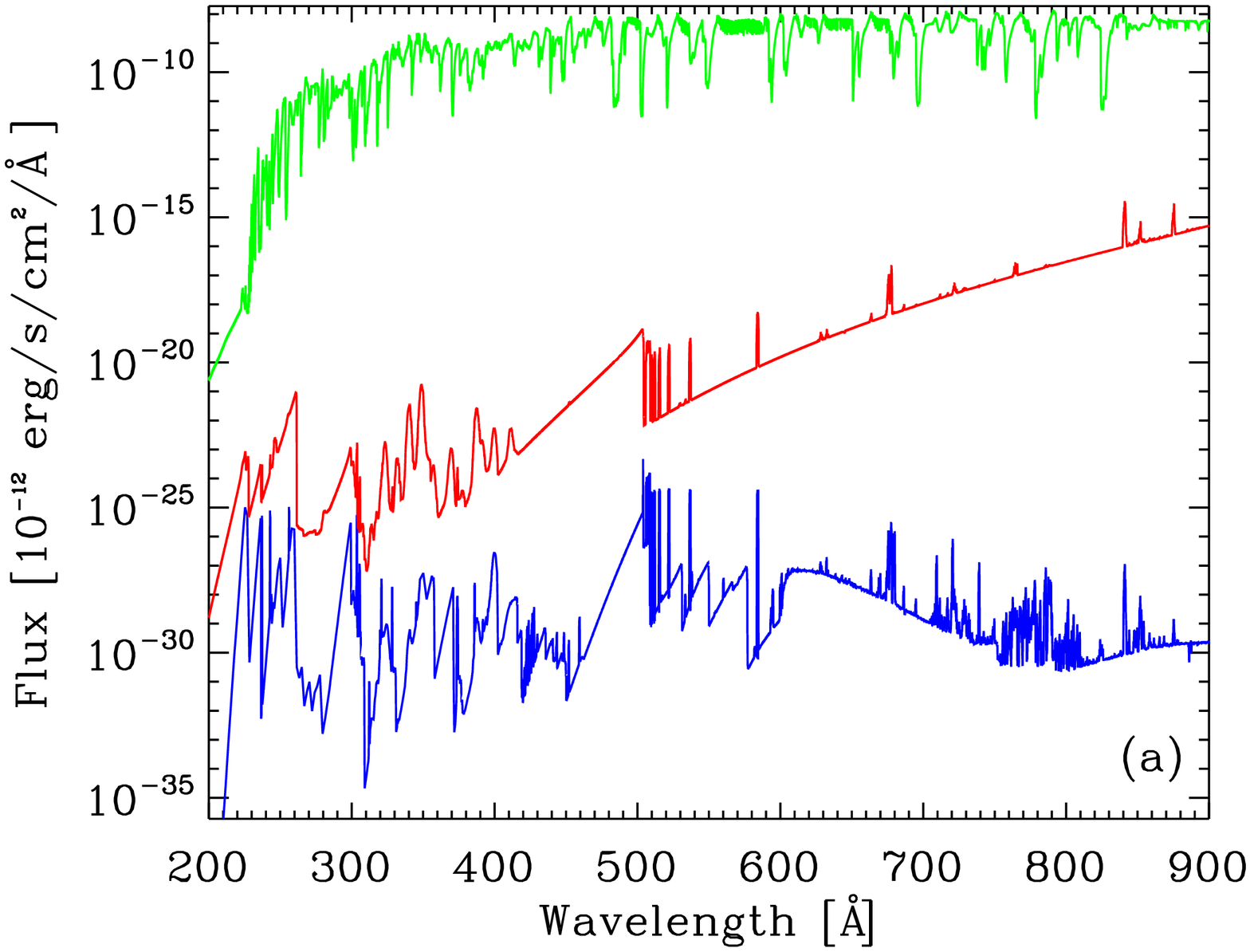}}\\
\resizebox{\hsize}{!}{\includegraphics{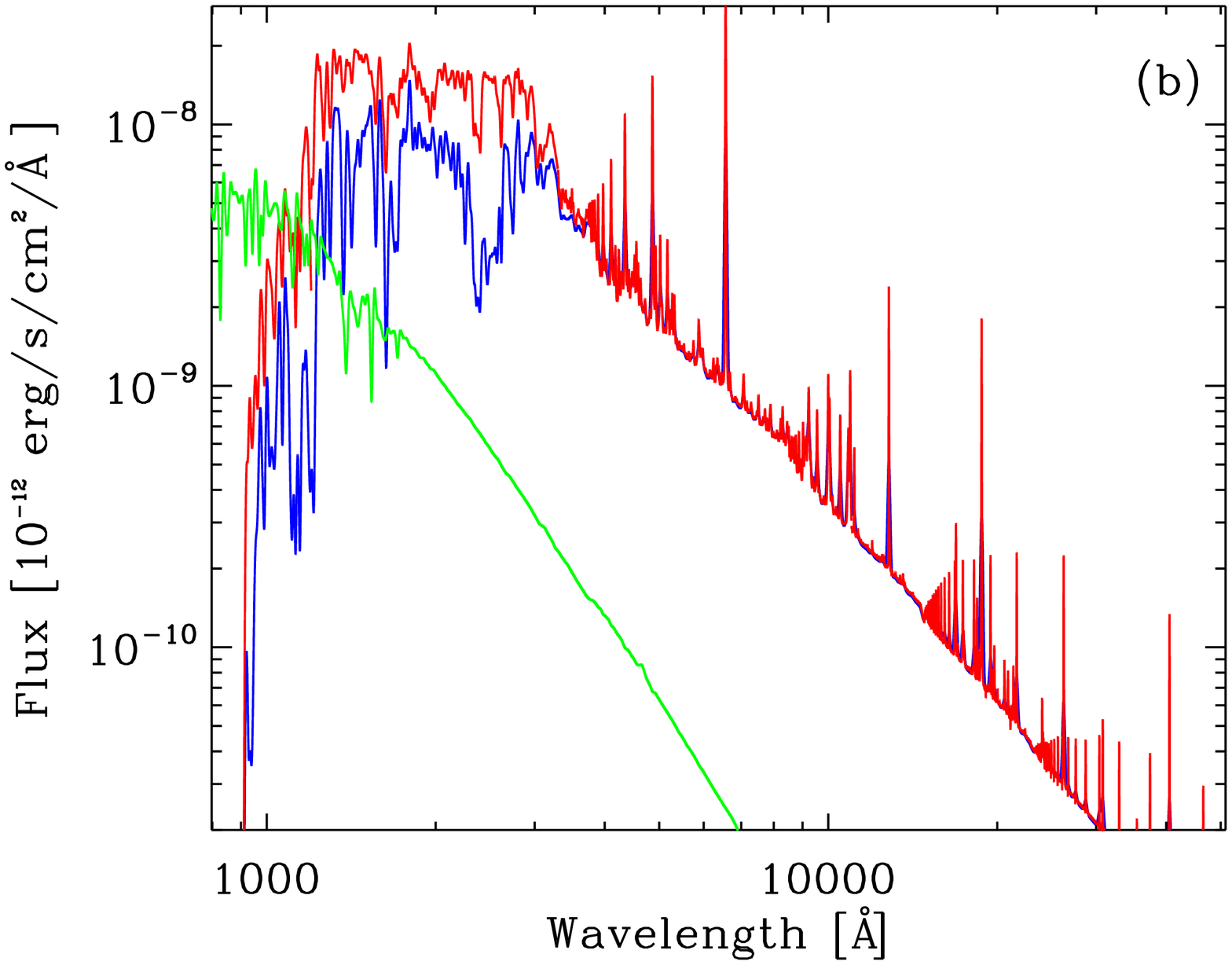}}\\
\resizebox{\hsize}{!}{\includegraphics{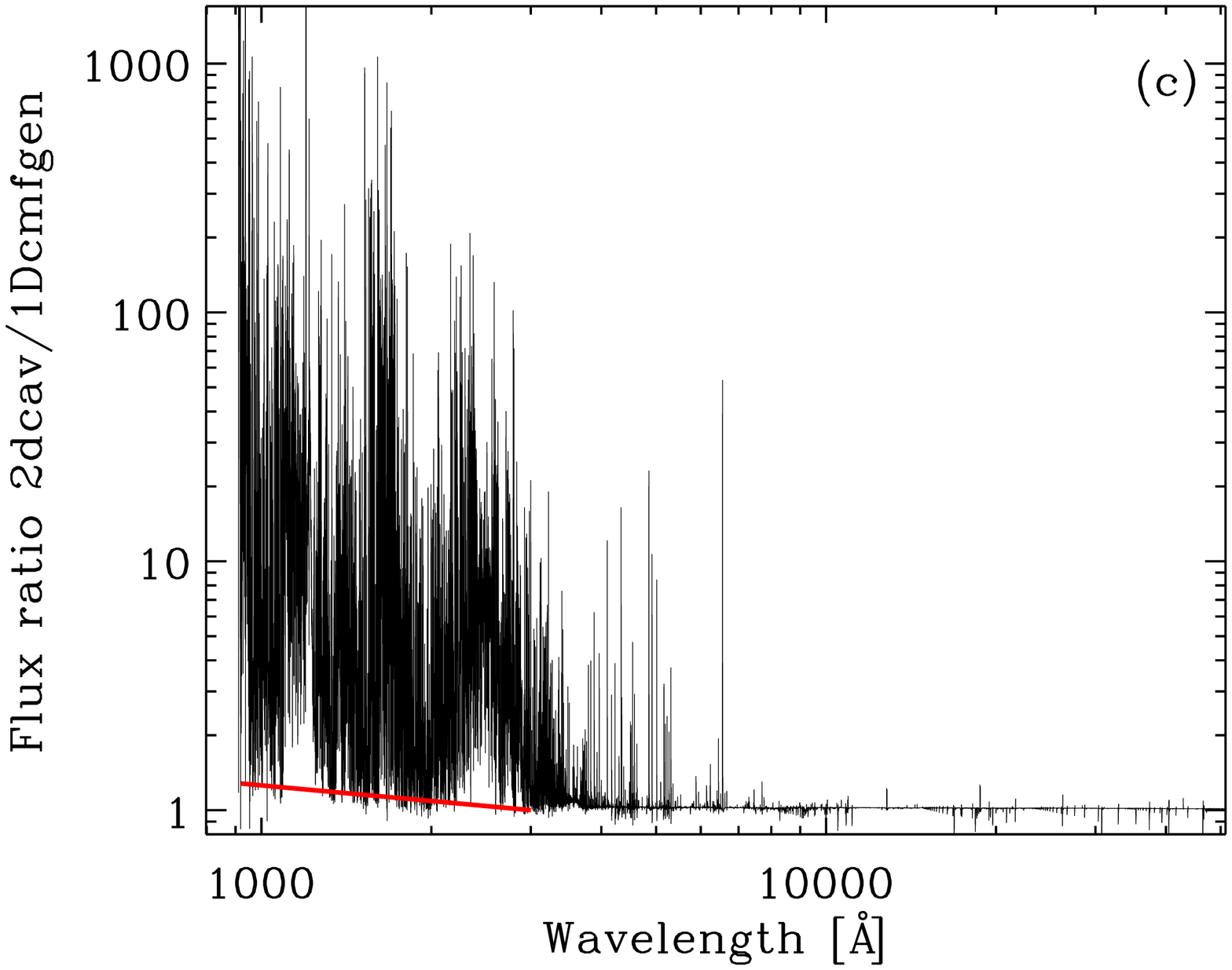}}
\caption{\label{sed}{{\it (a):} Spectral energy distribution in the range 200--900~\AA\ of our fiducial 2-D cavity model of \etaa\ assuming the parameters listed in Table \ref{params} (red line) and the 1-D CMFGEN model of \etaa\ (blue line). For comparison, the 1-D CMFGEN model of \etab\ is also shown (green line). {\it (b):} Idem, for the wavelength range  900-30000~\ang. All models were degraded to a spectral resolution of 1000 to help visualization in the UV and optical.{\it  (c):} Flux ratio between the 2-D cavity model spectrum and the 1-D CMFGEN model spectrum. The red line corresponds to the flux ratio inferred for monochromatic continuum regions free of \ion{Fe}{ii} lines.}}
\end{figure}

We now turn attention to the 2-D cavity models. Figures \ref{sed}a,b show the spectral energy distribution predicted by the 2-D cavity and the 1-D CMFGEN models from the far-UV to the near-IR. As anticipated from Fig. \ref{taulambda}, our 2-D modeling reveals that the spectral energy distribution (SED) of \etaa\ is strongly modified below 3000~\AA\ at apastron when the WWC cavity is open toward the observer (Figs. \ref{sed}a,b). 

Between 200--912~\ang, we find that a strong bore-hole effect may occur. The 2-D cavity model indicates that an observer viewing the binary system down the cavity toward \etaa\ would detect a flux from 5 to 10 orders of magnitude higher than that arising from the 1-D model lacking a cavity. However, even with this extreme increase in flux, a hot, luminous companion star would easily dwarf the emission from the primary star between 200--912~\AA\ (Fig. \ref{sed}a). Because of the steep radial dependence of the source function at these wavelengths, we anticipate that as the WWC cavity penetrates deeper and deeper at orbital phases closer to periastron, \etaa\ will contribute an increasing amount of flux between 200--912~\AA. Whether this increase is enough to rival the flux emitted by \etab\ and produce significant photoionization of the ejecta depends on detailed radiative transfer calculations and on the properties of \etaa, such as the assumed $\rstar$. 

From 3000~\AA\ up to $80~{\micron}$, we find that the 2-D cavity model at apastron has an SED very similar to the one from the 1-D CMFGEN model, illustrating that no bore-hole effect occurs around apastron in this wavelength range (Fig. \ref{sed}b). This is consistent with the findings from, e.\,g, \citet{madurathesis10} and T. I. Madura et al. (2012, in prep.).

Between 912--3000~\ang, however, there is a significant extra amount of emission in the 2-D cavity model compared to the 1-D CMFGEN model (Fig.~\ref{sed}b), even though the cavity does not penetrate into the optically-thick continuum regions of the wind of \etaa\ for these wavelengths (Fig.~\ref{taulambda}). We find that the increased continuum flux in the UV is caused by two effects: the modification of the continuum optical depth scale of \etaa's wind by the cavity, and the reduction in \ion{Fe}{ii} line blanketing. The latter effect clearly dominates the overall flux increase, causing huge flux ratios of around 10-1000 between 2-D and 1-D models within very narrow spectral regions (Fig. \ref{sed}c), corresponding to (blends of) spectral lines. The effect of the WWC cavity on spectral lines is discussed in Sect \ref{standardparamsline}. 

The increased continuum flux in the range 912--3000~\AA~ dilutes any putative emission lines from the companion, making it challenging to directly detect spectral signatures from \etab. In particular, beyond $\sim1500~\ang$, \etaa\ strongly dominates the flux, which has important consequences for the illumination of the circumbinary ejecta \citep{mg12}. We also note that the observed spectral energy distribution is different from the model predictions because of the anomalous reddening toward Eta Car \citepalias{hillier01}.

We further investigate why around apastron there is an increase in flux at $\it continuum$\footnote{I.\,e., regions not contaminated by \ion{Fe}{ii} lines.} wavelengths between 912--3000~\ang\ and not at longer wavelengths, such as in the near-IR. Recalling that the intensity at a given frequency $\nu$ and impact parameter $p$ of a ray with source function $S_\nu (\tau)$ is (Mihalas 1978)
\begin{equation}
\label{ipsph}
I_{p,\nu}=\int^{\infty}_0{S_\nu(\tau) e^{-\tau} d\tau}\, ,
\end{equation}
we study the behavior of $\tau$ and $S_\nu (\tau)$ in the UV and near-IR. For simplicity, we consider a ray with impact parameter $p=0$.

\begin{figure}
\center
\resizebox{\hsize}{!}{\includegraphics{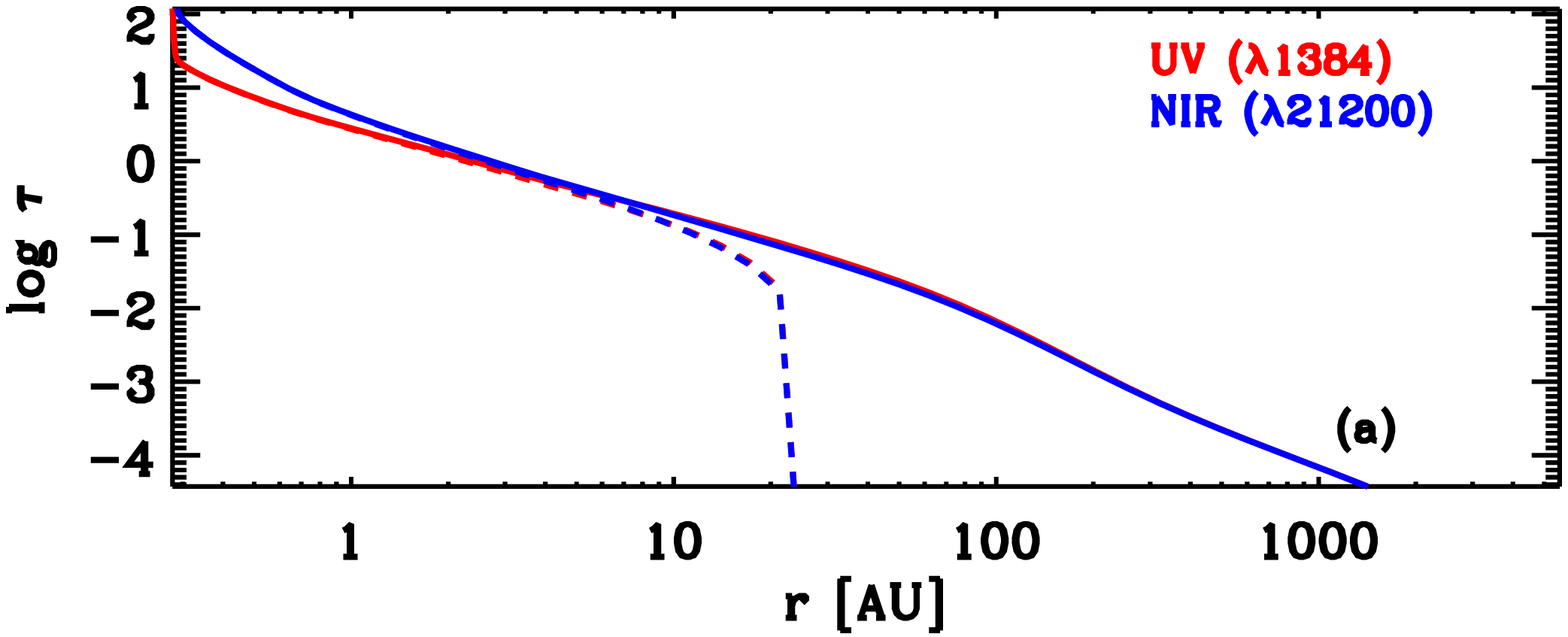}}
\resizebox{\hsize}{!}{\includegraphics{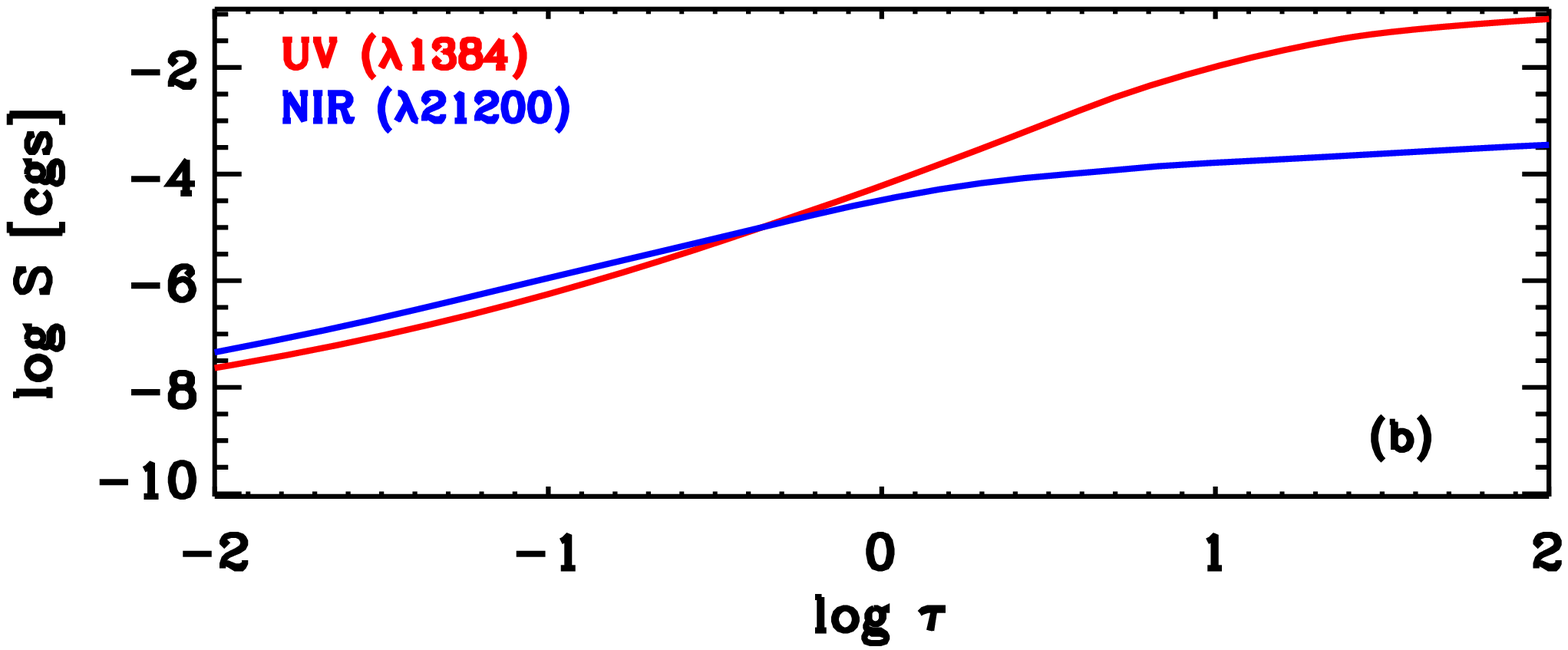}}
\resizebox{\hsize}{!}{\includegraphics{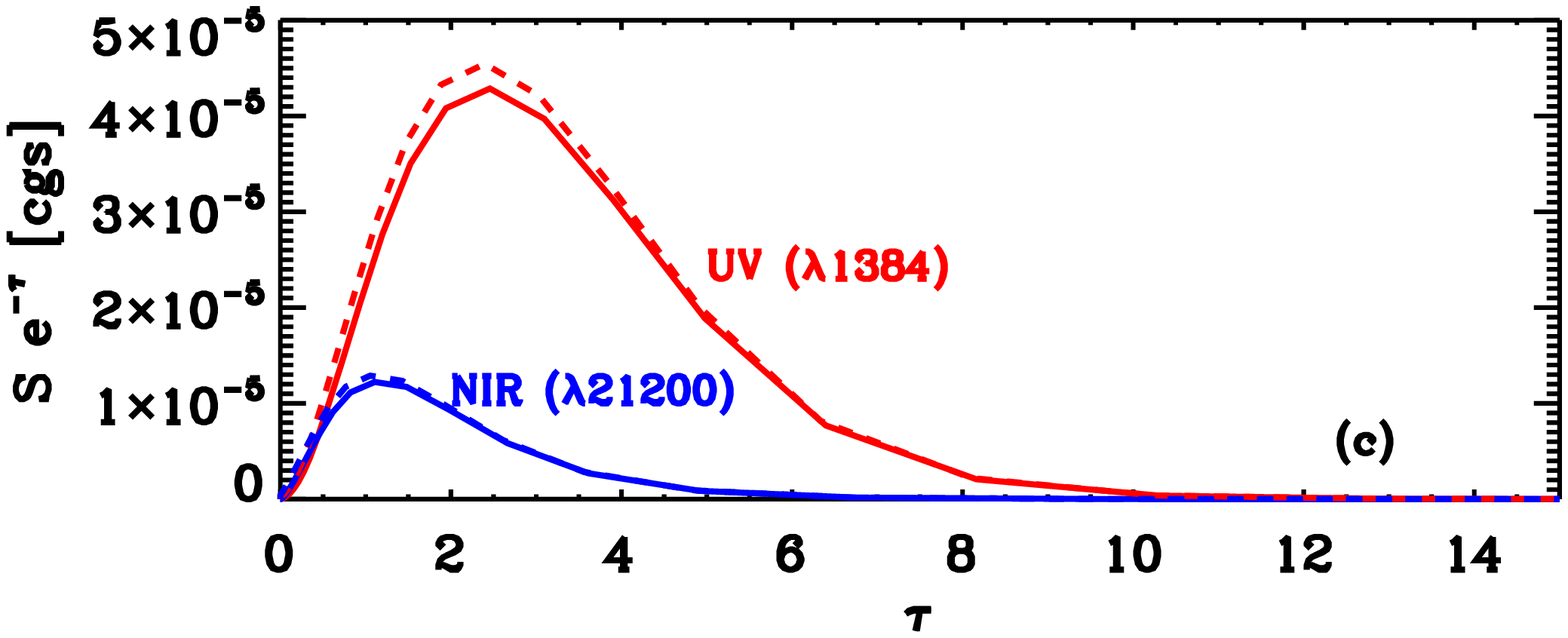}}
\caption{\label{tauchange}{{\it (a):} Radial variation of the continuum optical depth of a ray with $p=0$ (i.e., toward the core of \etaa) for the 1-D CMFGEN (solid lines)  and 2-D cavity models (dashed lines). We show representative wavelengths in the UV ($\lambda1384$, red lines) and near-IR ($\lambda21200$, blue lines). Note that the dashed and solid lines are superimposed for $r<2$~AU, as are the blue and red dashed lines for $r>8$~AU. {\it (b):} Variation of the continuum source function of \etaa\ in the UV ($\lambda1384$) and near-IR ($\lambda21200$) as a function of the respective continuum optical depth. {\it (c):} Idem, but for the intensity-like quantity $S e^{-\tau}$ as a function of $\tau$.}}
\end{figure}

Figure \ref{tauchange}a presents the variation of $\tau$ in the UV ($\lambda=1384~\ang$, red line)  and near-IR ($\lambda=21200~\ang,$ blue line) continua as a function of distance to \etaa\ for the 2-D cavity (dashed lines) and 1-D models (solid lines). We find that while the apex of the WWC cavity does not reach regions of the primary wind with $\tau > 2/3$ between $912-800000~\ang$, the presence of the WWC cavity still modifies the continuum optical depth of \etaa's wind at these wavelengths (Fig. \ref{tauchange}a). This occurs because material within the cavity has a much lower density, and thus much lower opacity $\chi$,  than that of the unmodified primary wind. Since in the usual $p-z$ geometry (Mihalas 1978)
\begin{equation}
\tau=\int^{z_{max}}_z{-\chi dz'}\, ,
\end{equation}
the reduction in $\chi$ implies a reduction in the optical depth at a given distance $z$ from the primary star. 

\begin{figure*}
\center
\resizebox{0.87\hsize}{!}{\includegraphics{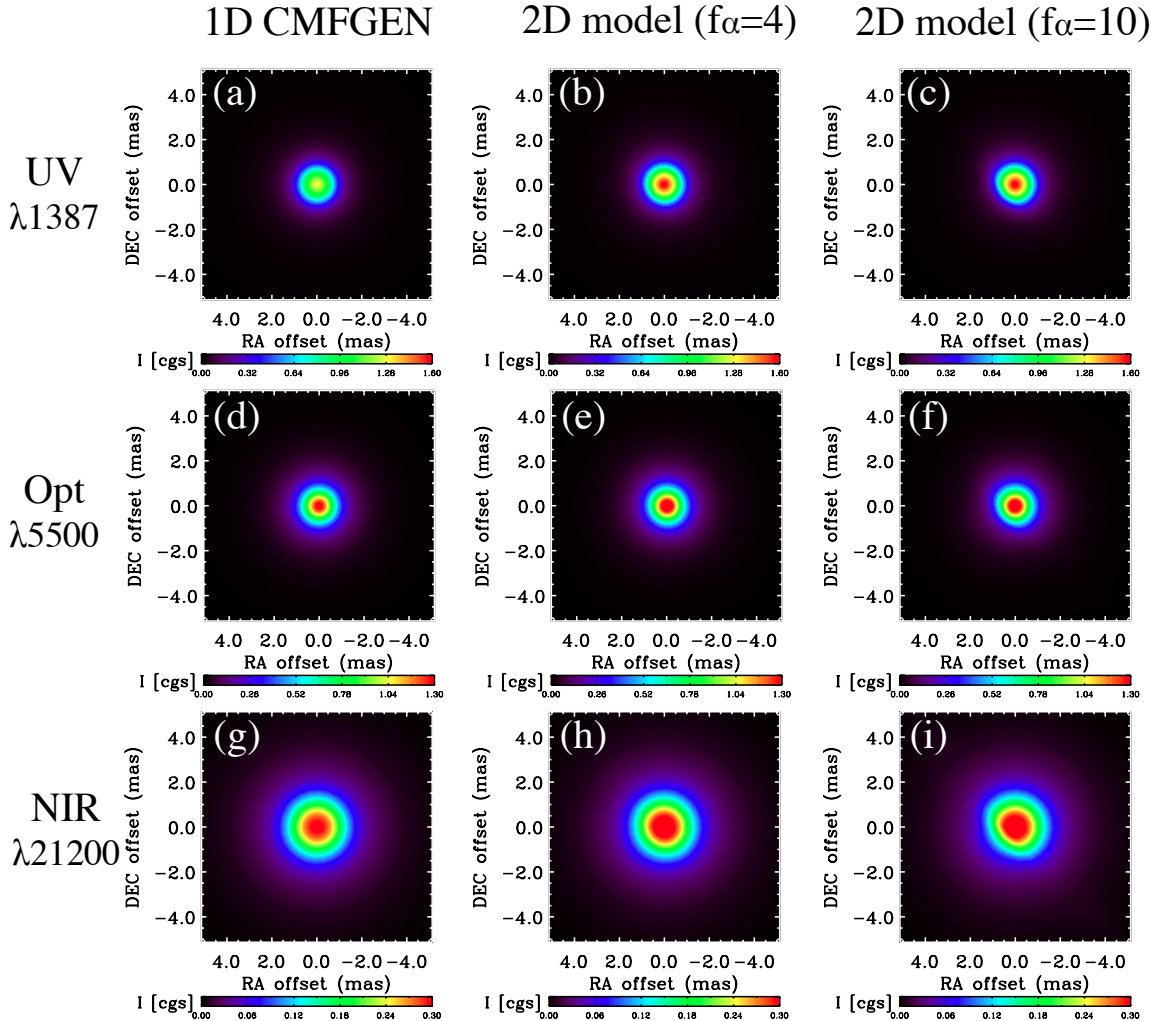}}
\caption{\label{ipuv}{Monochromatic continuum images predicted by the 1-D CMFGEN model (left column), 2-D cavity model with the parameters from Table \ref{params} (middle), and 2-D cavity model with a denser WWC zone with $f_\alpha=10$ (left). The top row shows the UV region at $\lambda1384$, the optical ($\lambda5500$) is presented in the middle row, while the near-IR region ($\lambda21200$) is displayed in the bottom row.}}
\end{figure*}

The variation of $S_\nu$ as a function of $\tau$ in the UV and near-IR is shown in Fig. \ref{tauchange}b. Our models indicate that $S_\nu (\tau)$ is steeper toward large $\tau$ at UV wavelengths than in the near-IR. This property explains the larger relative variation in the intensity-like quantity  $S_\nu(\tau) e^{-\tau}$ at $p=0$ in the UV compared to the near-IR (Fig. \ref{tauchange}c). Therefore, our 2-D models show that the change in the optical depth scale affects more the UV region than the optical/near-IR. Figure \ref{ipuv} presents monochromatic images in the UV, optical, and near-IR continua predicted by our 1-D and 2-D models, where the effects of the modification of the optical depth scale can be visualized. 

Using similar reasoning, the high value of  $S_\nu$ at large $\tau$ in the UV also causes the 2-D cavity models to have higher values of $I$ at large $p$ compared to the 1-D model (Figs. \ref{ipuv}b,c). Ultimately, this results in a higher flux in the UV, of the order of 20-30\% compared to the flux yielded by the spherical model (Figs. \ref{sed}). In the optical/near-IR, however, little changes are seen (Figs. \ref{ipuv}e,f,h,i). 

In addition, our 2-D model indicates that the presence of the WWC cavity affects the intensity distribution of the images (and hence the flux) via emission from and absorption by the shocked walls. As shown in \citet{gmo10} for the $K$-band continuum, these effects can potentially change the morphology of the images and the amount of flux arising at large impact parameters.

Emission and absorption from the WWC zone will be more noticeable in regions of high opacity/emissivity, such as in the UV (due to bound-free opacities from iron-group elements)  and infrared (due to free-free processes).  On one hand, the material in the WWC zone adds extra opacity in the line-of-sight to the continuum emitting region, which might decrease considerably the flux reaching the observer, depending on $f_\alpha$ (Fig~\ref{ipuv}). On the other hand, the presence of dense material in regions not covering the continuum source increases the emissivity of these regions. As a consequence, the amount of flux coming from large impact parameters will be higher. It is worth noting that emission and absorption from the WWC zone will be enhanced at orbital phases around periastron \citep{gmo10}, since the shock-compressed regions reach deeper into the wind of \etaa.

An important prediction of our 2-D models, which include the effects of \etab\ via the presence of a WWC cavity and WWC zone, is that the morphology of the continuum and spectral line emitting regions should be dependent on the orbital phase. As can be seen in Fig.~\ref{ipuv}, our models predict a roundish brightness distribution on the sky at apastron, with some asymmetry depending on the density and ionization structure of the walls of the WWC zone. Our models suggest that the image morphology should get increasingly more asymmetric and elongated at orbital phases closer to periastron \citep{gmo10}. This is in markedly contrast with the prediction from a single, rapid-rotator star scenario as proposed by \citet{smith03}, where rapid rotation during apastron should produce elongated, asymmetric images \citep{gmo10}.  In addition, no time dependence in the image morphologies should be seen, unless the rotational velocity of \etaa\ changes across the orbital cycle. Long-baseline optical interferometric observations are currently able to deliver model-independent, reconstructed images that would allow one to test the aforementioned scenarios.

\section{Ionization effects from the companion star} \label{ionization}

At present, the 2-D code from \citetalias{bh05} does not allow for the variation of the ionization stage and level populations as a function of latitude. Therefore, material inside the WWC cavity and along its shocked walls have the same ionization structure as the wind of \etaa. This is admittedly a zero-order approximation in the case of the Eta Car binary system, mainly because of the two factors elaborated below.

First, \etab\ supposedly has  $\teff=36,000-41,000~\K$ \citep{verner05,teodoro08,mehner10} and, depending on its luminosity, might be able to ionize hydrogen and other species in a non-negligible fraction of the wind of \etaa. This could affect the P-Cygni absorption of hydrogen lines, and give rise to spectral lines of high-ionization stages (such as \ion{Si}{iv}$\lambda\lambda~1398,1402$) that would otherwise not be present in the spectrum of \etaa\ in a single star scenario. \citet{madura11} estimated the volume of the wind of \etaa\ that is photoionized by \etab\ and found that a significant fraction of H could be ionized near apastron (see their Fig. 6). This result has been confirmed by \citet{kruip11} using independent radiative transfer modeling. However, as a caveat, both calculations include only H and He, while \citetalias{hillier01} showed that line blanketing by the Fe group elements is a key factor to the H ionization structure in dense stellar winds, as is the case of \etaa.

Second, material is shocked along the wall, significantly increasing its temperature, and is compressed and radiatively cools on the side of the primary. In addition, ionizing photons from \etab\ illuminate the post-shocked primary wind, and will affect this material. Hot, post-shocked secondary wind material will also be in line-of-sight and likely emitting in X-rays \citet{parkin09,parkin11}. Because of these complications we are unable to self-consistently model the line emission and absorption from the walls of the shock cone. However, they do not affect the conclusions reached in this paper regarding the WWC {\it cavity} effects. We note that the effects described above should affect much less the continuum emission and absorption, in particular the near-IR region studied in \citet{gmo10}.

As a consequence of these limitations, our model is not able to reproduce high-ionization lines thought to arise from material photoionized by \etab. We defer a detailed model of ionization effects to future work, when it will become feasible to couple 3-D radiative transfer tools (e.g., {\sc SIMPLEX}; \citealt{paar10,kruip11}) to 3-D hydrodynamical simulations. Nevertheless, despite the limitations, our 2-D models are adequate for understanding how the line profiles of \etaa\ are modified by the carving of its wind by \etab.

\section{Concluding remarks: implications of the WWC cavity} \label{conc}

We presented the first 2-D radiative transfer modeling of the central source in Eta Carinae, taking into account the carving of the stellar wind of the primary star that occurs due to the wind of a close companion star. We investigated the effects of the cavity using multi-wavelength diagnostics around apastron of the orbital cycle. We believe that the 2-D models presented here pave the way for future work on 3-D, full line blanketed radiative transfer modeling of the Eta Car binary system, that are warranted to gain further insights on the effects of \etab\ on \etaa.

The 2-D models have several  important implications for the understanding of the Eta Car binary system. Our models predict that the WWC cavity induces latitudinal and azimuthal dependence in the continuum and line profiles, which could be tested observationally with spectra reflected off the ejecta that surrounds the binary system. Such an investigation might provide clues on whether or not \etaa\ is a rapid rotator.

Our 2-D models unambiguously show that lines formed over a large volume of the wind of \etaa, such as H$\alpha$, H$\beta$, and \ion{Fe}{ii} lines, are affected by the cavity even during apastron. In addition, ionization from \etab\ may also affect the strength and shape of these lines. Therefore, analyses of long-term variations of Eta Car, possibly caused by changes in any stellar or wind parameter of one (or both) stars, should take into account the effects discussed in this paper.

In addition, the presence of a WWC cavity and WWC zone with an orientation and structure predicted by hydrodynamical simulations (e.\,g. \citealt{madura11}) causes the line-of-sight to the central source to have peculiar conditions. The combination of a half-opening angle of the WWC cavity  $\sim50\degr-60\degr$ and orbital inclination angle $\sim140\degr$ implies that the line-of-sight to the central source might cross a significant fraction of the post-shocked, dense regions of the WWC zone. As we discussed in this paper, emission and absorption of atomic species may occur. In addition, if dust is able to form along the WWC zone, as suggested from studies of WR binaries \citep[e.\,g.][]{williams09,williams11} and Eta Car itself \citep{diego05,kashi08a,smith10}, emission from the central source would suffer from dust extinction. This scenario might explain the anomalous extinction seen towards the central source compared to ejecta just $\sim0\farcs3$ away \citep{hillier92}, perhaps explaining the ``coronagraph" effect discussed by \citetalias{hillier06} \citep{smith10}. Moreover, Eta Car has displayed a puzzling photometric brightening in the last century \citep[e.\,g.][]{frew04}, and long-term changes in the opening angle of the WWC cavity and/or dust formation rate along the WWC zone might be one of the culprits.

Finally, we believe the 2-D models presented here provide the foundation to understand the behavior of spectral lines at phases around periastron, even if 3-D models are ultimately needed because of the wrapping of the WWC cavity around \etaa. Our models predict that the amount of P-Cygni absorption seen in H and \ion{Fe}{ii} lines depends on the amount of wind of \etaa in the line-of-sight to the observer. Since 3-D hydrodynamical simulations show an increase in material from  \etaa\'s wind in the line-of-sight just after periastron \citep{okazaki08,parkin09,madura11}, we suggest that the increase in the P-Cygni absorption of H and \ion{Fe}{ii} lines observed at these phases \citep{damineli98,nielsen07,richardson10} is caused  by the higher optical depth of these lines due to the increased amount of \etaa's wind in the line-of-sight to \etaa.

Below we summarize our main findings and conclusions.
\begin{list}{\arabic{qcounter}.~}{\usecounter{qcounter}}
\item We made  a detailed comparison of observed \hststis\ spectra near apastron ($\phisp=10.410$ and 10.603) with 1-D spherical CMFGEN radiative transfer models. We find that a revised wind terminal velocity of $\vinf=420~\kms$ provides better agreement with the redshifted part of H and \ion{Fe}{ii} emission lines in the optical spectrum, and the P-Cygni absorption component of low-ionization UV resonance lines of \ion{C}{ii}, \ion{Si}{ii}, \ion{Fe}{ii}, and \ion{Mg}{ii}. We find the following parameters for our revised spherical CMFGEN model of \etaa: $\vinf = 420~\kms$, $\mdot = 8.5 \times 10^{-4}~\msunyr$, $f = 0.1$, \tstar~=~35200~K (at $\tau_\mathrm{Ross}=130$), \teff = 9400~K (at $\tau_\mathrm{Ross} = 2/3$),  $\lstar = 5 \times 10^6~\lsun$, assuming a distance of $d = 2.3$ kpc. 

\item We adapted the 2-D radiative transfer code of \citetalias{bh05} to account for the modification of the wind of \etaa\ caused by the presence of a low-density WWC cavity and a dense WWC zone. We investigated in detail the effects of the WWC cavity on spectral lines, showing that two factors dominate the influence: the size of the line formation region compared to the distance of the apex of the WWC cavity, and the orientation of the cavity with respect to the observer. For our preferred orbital orientation with $i=138\degr$ and $\omega=260\degr$, the WWC cavity affects more the blueshifted and zero velocity parts of spectral line profiles.

\item According to our 2-D radiative transfer modeling, H$\alpha$, H$\beta$, and \ion{Fe}{ii} lines are the most affected by the WWC cavity, since they form over a large volume of \etaa's stellar wind.  The excess P-Cygni absorption seen in H$\alpha$, H$\beta$, and optical \ion{Fe}{ii}  in spherical models becomes much weaker or absent in 2-D cavity models with  $i=138\degr$ and $\omega=260\degr$, in agreement with the \hststis\ observations around apastron.

\item  The same 2-D model simultaneously provides a superb fit to the observed UV spectrum of Eta Car at apastron. The better fits are because the UV spectrum of Eta Car is strongly dominated by absorption of \ion{Fe}{ii} lines that become significantly weaker when the presence of the low-density WWC cavity is taken into account.

\item We obtain that orbital orientations with $110\degr \la i \la  140\degr $ and $210\degr \la \omega \la 330\degr$  are needed to simultaneously fit the UV and optical spectrum of Eta Car around apastron, assuming $\alpha$ in the range $50\degr-70\degr$. Larger values of $\alpha$ and/or photoionization of the wind of \etaa\ and WWC zone by \etab\ could change the range of allowed $i$ and $\omega$ slightly.

\item  Our models predict that the WWC cavity induces latitudinal and azimuthal dependence in the continuum and line profiles. In particular, the P-Cygni absorption is stronger in spectral lines viewed from $i=180\degr$ than at the central star position. Therefore, our models predict that spectra scattered off the Homunculus poles should have stronger P-Cygni absorption than those observed on the central star or equatorial positions. The presence of the WWC cavity thus mimics the effects that could, in principle, also arise in a single, fast-rotating \etaa.

\item Our 2-D model indicates that the apex of the WWC cavity does not reach optically-thick regions in the continuum in the UV, optical, and infrared during apastron, and that these regions will be affected by a bore-hole effect only at  $0.95 \la \phiorb \la 1.05$. This is in agreement with the findings from \citet{madura10,madurathesis10,madura11a}.

\item We find that for continuum wavelengths with $\lambda \la 912~\ang$ or $\lambda \ga80~{\micron}$, the WWC cavity penetrates the $\tau=2/3$ surface at apastron. For $\lambda \la 912~\ang$ a pronounced bore-hole effect occurs, increasing by 5 to 10 orders of magnitude the number of photons escaping through the WWC cavity. However, this extra flux is easily dwarfed by that expected from the hot, luminous companion \etab\ during apastron.

\item For $912 \la \lambda \la 2000~\ang$, even if the WWC cavity  does not penetrate the $\tau=2/3$ surface at apastron, a flux increase of the order of 10--30\% is seen in the 2-D model with a WWC cavity open toward the observer compared to the spherical 1-D model. We found that the flux increase is not only caused by the modification of the optical depth scale of the primary wind, but also by the steeper optical depth dependence of the source function in the UV compared to the optical/near-IR. 

\item We present predictions of continuum images at apastron in the UV, optical, and near-IR, that could be tested observationally with interferometric techniques. In addition to the bore-hole effect, emission and absorption from the dense walls of the WWC zone might significantly affect the continuum images and flux, depending on the orbital orientation and density of the walls.

\end{list}

\section*{Acknowledgements}

We thank our referee, Dr. A. F. J. Moffat, for the careful reading and detailed comments. We gratefully acknowledge many useful discussions with M. Corcoran, A. Damineli,  T. R. Gull, S. Owocki, and N. Smith. JHG and TIM thank the Max Planck Society for financial support. D. J. Hillier acknowledges support from HST-GO-11623.01-A and HST-GO-11756.01.This research made use of the Smithsonian NASA/ADS and SIMBAD (CDS/Strasbourg) databases.

\def\aj{{AJ}}                   
\def\araa{{ARA\&A}}             
\def\apj{{ApJ}}                 
\def\apjl{{ApJ}}                
\def\apjs{{ApJS}}               
\def\apss{{Ap\&SS}}             
\def\aap{{A\&A}}                
\def\aapr{{A\&A~Rev.}}          
\def\aaps{{A\&AS}}              
\def\mnras{{MNRAS}}             
\def\pasp{{PASP}}               
\def\solphys{{Sol.~Phys.}}      
\def\sovast{{Soviet~Ast.}}      
\def\ssr{{Space~Sci.~Rev.}}     
\def\nat{{Nature}}              
\def\iaucirc{{IAU~Circ.}}       

\let\astap=\aap
\let\apjlett=\apjl
\let\apjsupp=\apjs
\let\applopt=\ao

\bsp

\label{lastpage}

\end{document}